\documentclass[fleqn,usenatbib]{mnras}

\usepackage{newtxtext,newtxmath}

\usepackage[T1]{fontenc}

\DeclareRobustCommand{\VAN}[3]{#2}
\let\VANthebibliography\thebibliography
\def\thebibliography{\DeclareRobustCommand{\VAN}[3]{##3}\VANthebibliography}

\usepackage{graphicx}	
\usepackage{amsmath}	
\usepackage{bm}
\usepackage{setspace}

\makeatletter
\newlength{\abovecaptionskip}%
\makeatother
\usepackage[flushleft]{threeparttable}

\usepackage{array}

\defcitealias{2005ApJ...623..398Y}{Y05}

\title[Nova remnants in the GC]{Hydrodynamic Evolution and Detectability of Nova Remnants in the Galactic Center}

\author[Z. Su \& Z. Li]{
Zhao Su $^{1,2}$\thanks{E-mail: suzhao@smail.nju.edu.cn (ZS)} and 
Zhiyuan Li$^{1,2,3}$\thanks{E-mail: lizy@nju.edu.cn (ZL)}
\\
$^{1}$School of Astronomy and Space Science, Nanjing University, Nanjing 210046, China\\
$^{2}$Key Laboratory of Modern Astronomy and Astrophysics (Nanjing University), Ministry of Education, Nanjing 210046, China\\
$^{3}$Institute of Science and Technology for Deep Space Exploration, Suzhou Campus, Nanjing University, Suzhou 215163, China
}

\date{Accepted XXX. Received YYY; in original form ZZZ}

\pubyear{2025}

\begin{document}
\label{firstpage}
\pagerange{\pageref{firstpage}--\pageref{lastpage}}
\maketitle

\begin{abstract}
Thousands of X-ray sources have been detected in the Galactic center (GC), most believed to be cataclysmic variables (CVs).
As a potential probe of the old stellar population, in particular CVs, the existence and detectability of novae in the GC remain elusive, due to the prohibitive extinction toward the GC and their relatively low occurrence rate.
Nova remnants evolving in the characteristic hot ($T\sim{10^{6}~\rm K}$) and dense ($n_e\sim{10~\rm cm^{-3}}$) interstellar medium in the GC may shed light on recent novae and provide useful insight on the GC ecosystem.
In this work, we perform hydrodynamical simulations of putative nova remnants in the GC environment and calculate their time-dependent multiwavelength emission to estimate the detectability.
Among 79 models sampling the nova parameter space (primarily ejecta mass and velocity), 6, 44, and 51 modelled nova remnants are detectable at their X-ray, radio, and Paschen-$\alpha$ maximum, respectively, for existing {\it Chandra}, VLA, and HST observations of the GC.
The predicted peak luminosities are $\sim10^{32}~\rm erg~s^{-1}$, $\sim10^{31}~\rm erg~s^{-1}$, and $\sim10^{36}~\rm erg~s^{-1}$ in these three bands and the detectable window ranges from weeks to notably hundred years.
By specifying a CV population of the nuclear star cluster, we estimate the probability of detecting at least one remnant to be 20\%, 8\%, and 18\% in X-rays, radio, and Pa$\alpha$. 
The nova remnant would be best resolved in the X-ray band. 
Our study highlights the potential for detecting nova remnants through further observations, leveraging JWST and the potentially forthcoming AXIS and SKA.
\end{abstract}




\section{Introduction}\label{sec:intro}
Nova is among the most common thermonuclear explosions in the universe, as a result of unstable nuclear burning on the surface of an accreting white dwarf (WD) (see \citealp{2021ARA&A..59..391C} for a recent review).
Accreted material from the companion star mixed with the WD matter is ejected during the outburst.
The ejecta covers a wide span in total mass of $\sim10^{-7}-10^{-3}~\rm M_\odot$ \citep{2008clno.book.....B} and with expansion velocities in a range from $\sim10^2$ to several $10^3~\rm km~s^{-1}$ \citep{2018MNRAS.476.4162O}.
Conventionally, novae are categorized as recurrent or classical novae according to whether at least two outbursts have been observed.

The Galactic center (GC), particularly the nuclear star cluster, hosts a high density of cataclysmic variables (CVs) which are manifested as point-like X-ray sources \citep{2018ApJS..235...26Z}. 
The CV origin of these X-ray sources is supported by the facts that (i) the surface number density of the sources broadly follows the starlight distribution, and the X-ray source abundance is roughly consistent with the CV abundance in the solar neighborhood; and (ii) the cumulative X-ray spectra, exhibiting hard bremsstrahlung continuum and strong iron lines, are typical of CVs (see \citealt{2018ApJS..235...26Z} and references therein for details).
While this CV population is expected to be a nova factory, no nova bursts have been conclusively detected thus far.
The reason for this non-detection can be twofold: (i) the occurrence rate of novae is moderate, with $\approx20-100$ yr$^{-1}$ for the Milky Way depending on the measurement method \citep[e.g.,][]{2017ApJ...834..196S, 2021ApJ...912...19D, 2022ApJ...937...64K, 2023MNRAS.523.3555Z}; (ii) the severe obscuration toward the GC inhibits conventional optical monitoring.
In the context of {\it XMM-Newton} monitoring of the GC, \citet{2008ApJ...677.1248M} estimate the nova rate in the inner $15\arcmin$ region to be $\sim0.1~\rm yr^{-1}$ and suggest that a fraction of the X-ray transient sources could be unidentified novae.
When adopting updated measurements on nova rate of the Milky Way ($\approx40~\rm yr^{-1}$) and enclosed stellar mass in the GC ($\approx 10^8~\rm M_\odot$), the nova rate in this region is estimated to be $0.08~\rm yr^{-1}$ which is consistent with that by \citet{2008ApJ...677.1248M}.

Nova is a potential probe of CV population in the nuclear stellar cluster (NSC) as the occurrence rate and physical properties of novae are highly dependent on factors such as the WD mass, the mass transfer rate, and the chemical composition of the WD. \citep[e.g.,][]{2005ApJ...628..395T, 2015MNRAS.446.1924H, 2016ApJ...819..168H, 2016MNRAS.458.2916C, 2019MNRAS.490.1678C, 2018ApJ...860..110S, 2020A&A...634A...5J, 2020ApJ...895...70S, 2024ApJ...962..191S}.
With the rapidly expanding ejecta, nova would also help sweep out the ISM in the nuclear star cluster and potentially influence the accretion flow onto the central supermassive black hole, Sgr A*.
Additionally, novae might be the origin of unknown objects observed in the GC.
For example, one scenario for the enigmatic G2 object, a gaseous and dusty cloud in the vicinity of Sgr~A* \citep{2012Natur.481...51G}, is the ejected material from a recent nova outburst \citep{2012A&A...546L...2M}.

The interaction between nova ejecta and the interstellar medium (ISM) in the GC can generate strong shocks. 
The shocked ISM in nova remnants may unveil otherwise unnoticed novae in a time-scale much longer than the eruption stage typically lasting within a few weeks \citep{2008ApJ...677.1248M}, which is the focus of this work.
The expanding shell also provides a useful means of studying the physical conditions of the ISM (e.g., metal abundances) in the GC. 
In X-rays, four spatially extended remnants have been found surrounding known classical or recurrent novae, with shell radii of $\lesssim 0.1~\rm pc$, equivalent to an angular size of 21 arcsec at a distance of 1 kpc (see \citealp{2008clno.book.....B}; \citealp{2025Univ...11..105B} and references therein).
Their X-ray spectra favor that the X-ray emission is dominated by shock-heated gas, which indicates the interaction between the nova ejecta and the circumstellar medium \citep[e.g.,][]{2005ApJ...627..933B,2010MNRAS.404L..26B,2014A&A...572A.114B, 2022ApJ...926..100M}.
In addition, \citet{2006AdSpR..38.2840B} conducted an archival search for extended X-ray emission associated with known novae, but no newer X-ray remnants were discovered.
The main reason for the non-detection is that most novae were only observed by the ROSAT survey with a shallow exposure and a relatively low angular resolution of $\sim30~\rm arcsec$.
With the requirement for depth and angular resolution, the NSC would be a promising target since it was observed by {\it Chandra} in ultra-deep exposures with an unprecedented arcsec-resolution.
Moreover, with its extreme compactness, the NSC has probably the highest CV surface density in the Milky Way of up to $\sim10^{-2}~\rm arcsec^{-2}$ \citep{2018ApJS..235...26Z}.

The search for nova remnants in the NSC is further motivated by the expectation that the NSC may have an enhanced nova rate compared to the Galactic field.
On one hand, as a compact, massive stellar system, the NSC likely hosts an overabundant population of CVs due to dynamical effects.
\citet{2005ApJ...622L.113M} and \citet{2018ApJS..235...26Z} found an overabundance of transient X-ray sources within the inner parsecs of the Galactic center, indicating the role of dynamical process on X-ray binaries.
On the other hand, an enhanced nova rate by a factor of two is found in the region near the jet of M87 \citep{2024ApJ...973..144L}.
Although the origin of this jet-nova relation is still unclear, the putative jet launched by Sgr A*, possibly unveiled by the X-ray parsec-scale filament \citep{2013ApJ...779..154L,2019ApJ...875...44Z}, could also enhance the nova production in the NSC.

The evolution of supernova remnants (SNRs) in the GC has been extensively investigated.
This includes works concerning the specific SNR Sgr A East and its interaction with the nuclear outflow launched by massive stars in the central parsec \citep{2022ApJ...927L...6Z}, the generic morphological evolution of SNRs against a putative nuclear outflow \citep{2017ApJ...838...12Y}, and the effect of SNR expansion on the accretion onto Sgr A* \citep{2020A&A...644A..72P,2022MNRAS.510.5266B}.
However, the evolution and detectability of nova remnants which can be considered miniature analogs of SNRs, have not yet been systematically studied.

In this work, we investigate the hydrodynamic evolution and detectability of nova remnants in the Galactic center ($d\approx8~{\rm kpc}$, $1\arcsec$ corresponds to 0.039 pc; \citealp{2019A&A...625L..10G}, \citealp{2019Sci...365..664D}) by long-term, multi-wavelength observations. We have structured the paper as follows. In Section \ref{sec:hd}, we present hydrodynamical simulations of nova remnant evolution in the GC environment. In Section \ref{sec:detect}, we evaluate the multiwavelength detectability of the simulated nova remnants, especially in X-rays. In Section \ref{sec:prob}, we estimate the probability of nova remnant detection in X-ray, radio, and infrared band. We address further implications in Section \ref{sec:discussion} and summarize our study in Section \ref{sec:summary}.

\section{Numerical Simulation}\label{sec:hd}
To investigate the evolution of nova remnants in the GC and predict their detectability, we perform one-dimensional (1D) hydrodynamical simulations using the publicly available code {\sc pluto} \citep{2007ApJS..170..228M} with selected physical parameters for nova outbursts.
We adopt a 1D spherical grid with 20480 cells uniformly ranging from $2.4\times10^{-4}$ pc to 1.0 pc along the radial axis.
The ejecta is assumed to be isotropic by default.
We note that nova ejecta mass loss and velocity distribution can be anisotropic due to equatorial winds, bipolar ejections, and/or jets \citep[e.g.,][]{2003MNRAS.344.1219H,2008ApJ...688..559R,2009AJ....138.1541M,2011MNRAS.412.1701R,2020MNRAS.495.4372T}.
Additionally, our simulations do not account for clumpy ejecta with a low filling factor ($f\approx0.1$), as reported in some novae \citep[e.g.,][]{2018ApJ...853...27M}.
In this study, we aim to explore the basic evolution of nova remnants in the GC, while precise modelling of anisotropic and clumpy effects lies beyond the scope of this work.

The hydrodynamical equation is solved with a second-order Runge–Kutta time integrator and a Harten-Lax-van Leer Riemann solver for middle contact discontinuities (HLLC).
The simulations adopt an ideal equation of state with an adiabatic index of 5/3 and include optically thin radiative cooling for non-equilibrium ionization that is detailed in Section \ref{sec:photoionization}.
All simulations are run for 1000 yr after nova outburst. 
This duration is justified by the age of the observed nova remnants of up to several hundred years \citep[e.g.,][]{2020A&A...641A.122T} and by the extended radiative cooling timescales expected for novae evolving in the hot ISM.
Gravity and magnetic field are ignored in the simulations.
The numerical convergence is discussed in Appendix \ref{sec:resolution}.

\subsection{Initial \& Boundary condition}\label{subsec:init}
We assume a static, initial ISM having temperature $T=10^6~\rm K$ and electron number density $n_e=10~\rm cm^{-3}$. 
This assumption is consistent with {\it Chandra} observations of diffuse X-ray emission in central parsec in the GC \citep{2003ApJ...591..891B} and also favored by analytical and numerical wind-fed model for hot gas at a distance from $\sim1~\rm pc$ to the GC \citep{2004ApJ...613..322Q,2018MNRAS.478.3544R}.
The mass of the ISM enclosed in the simulation box is $0.15~\rm M_\odot$ that is at least two orders of magnitude higher than nova ejecta mass thus guarantee a complete cover of remnant evolution.
To emphasize the role of the hot ISM on the remnant evolution, we also neglect the circumstellar medium from putative common envelope evolutionary phase and the slow ejection component before the nova outburst.

The bursting WD is located at $r=0$. Due to the limitation of resolution, the simulation grid cannot resolve surface of the WD at $r_{\rm WD}\sim10^9~\rm cm$ \citep{2000A&A...353..970P}, and the ejecta is injected from inner boundary of the simulation grid at $r_{\rm in}\sim 10^6~r_{\rm WD}$. The ISM within the inner boundary has a mass of $10^{-11}~\rm M_\odot$ that is negligible relative to the ejecta mass thus would have little influence on the evolution. The ejection is mimicked by a continuous wind with constant velocity and the mass-loss rate transits from nearly constant to $\dot{M}\propto t^{-2}$ in a timescale $t_{\rm ej}$ \citep[see similar prescription in][]{2014MNRAS.442..713M}. Suppose the ejecta has total mass $M_{\rm ej}$ and velocity $v_{\rm ej}$, then at the inner boundary $r_{\rm in}$, the density at time $t$ relative to start of the eruption is	
\begin{eqnarray}
\rho &=&\frac{\dot{M}_{\rm ej}}{4\pi r_{\rm in}^2v_{\rm ej}} \nonumber\\
&=&\frac{1}{4\pi r_{\rm in}^2v_{\rm ej}}\frac{M_{\rm ej}}{\sqrt{\pi}t_{\rm ej}}\Big(1-e^{-(t/t_{\rm ej})^2}\Big)\Big(\frac{t}{t_{\rm ej}}\Big)^{-2}.
\end{eqnarray}
We note that the mass injection lasts for the entire simulation duration (1000 years), while a realistic nova outburst could have a much shorter time scale.
 However, the injection rate in the simulation approximately follows $\dot{M}\propto t^{-2}$ and will decrease to a negligible level within $\sim10t_{\rm ej}$, corresponding to a timescale of months to years.
 We also assess the impact of this injection prescription by turning off the injection one year after the nova outburst, and the simulation yield very similar results.

The temperature of the ejecta at the inner boundary is influenced by a combinative effect of adiabatic expansion and photoionization heating from the central WD in the super soft X-ray phase, which is described in Section \ref{sec:photoionization}.

\subsection{Parameter space}\label{subsec:para}
\begin{figure}
	\centering
	\includegraphics[width=\linewidth]{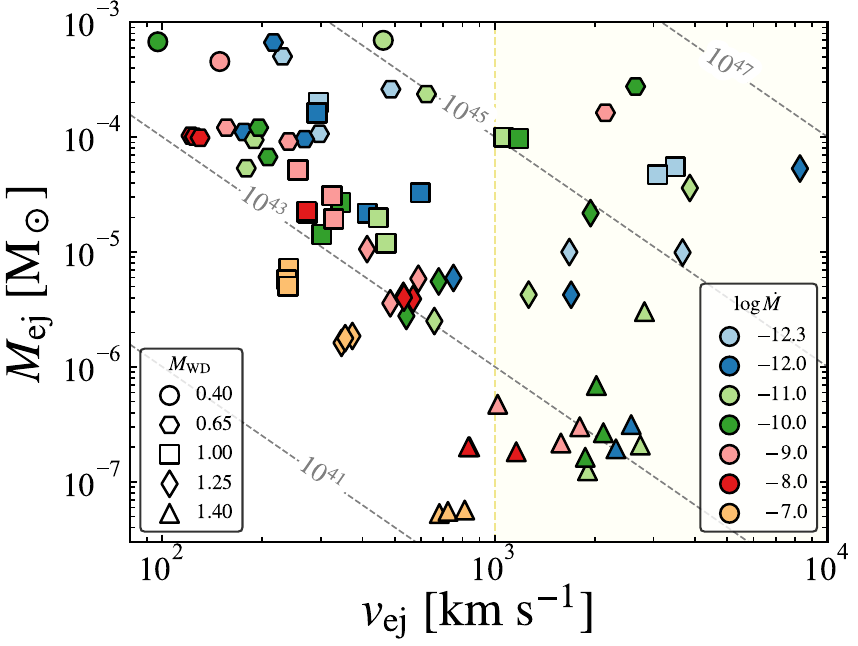}
	\caption{Ejecta masses and average velocities of simulated nova outbursts from \citetalias{2005ApJ...623..398Y} model grid. $M_{\rm WD}$ is WD mass in units of $\rm M_\odot$, differentiated by symbol shapes; $\log\dot{M}$ is mass transfer rate in units of $\rm M_\odot~yr^{-1}$ and logarithm, differentiated by symbol colors. The grey dashed lines indicate the kinetic energy of each nova outburst in units of erg.\label{fig:sample}}
\end{figure}

As for physical properties of nova ejecta, we adopt the extended model grid of nova outbursts from \citet[][hereafter Y05]{2005ApJ...623..398Y}.
The \citetalias{2005ApJ...623..398Y} model grid maps three key properties of accreting WD that control the outcome of a nova outburst: the mass of the accreting WD, $M_{\rm WD}$, the mass transfer rate, $\dot{M}$, and the core temperature, $T_{\rm WD}$.
The model grid consists of 99 parameter combinations: five in $M_{\rm WD}$ values ($0.4-1.4~\rm M_\odot$), eight in $\dot{M}$ values ($10^{-12.3}-10^{-6}~\rm M_\odot~yr^{-1}$), and three in $T_{\rm WD}$ values ($10-50~\rm MK$).
We perform simulations for 79 models (Figure. \ref{fig:sample}) with effective ejection, i.e., $v_{\rm ej}>0$.
As the GC is fully obscured from the optical to soft X-rays, hard X-rays ($\gtrsim 2~\rm keV$) potentially provide an effective window to detect nova remnants by tracing the shocked gas.  
For a strong shock, the relation between characteristic temperature and shock velocity is, 
\begin{equation}
	kT_{\rm sh}=\frac{3}{16}\mu m_p v_{\rm sh}^2\approx1.2\Big(\frac{v_{\rm sh}}{1000~{\rm km~s^{-1}}}\Big)^2~\rm keV.
\end{equation}
Among the 79 nova models, 25 exhibit an average ejecta velocity exceeding $1000~\rm km~s^{-1}$, suggesting that their remnants are likely to produce significant hard X-ray emission..

The ejecta mass, ejection time duration, and average ejecta velocity of each model are then transferred to the boundary condition described in Section \ref{subsec:init}.
If the recurrence time of the nova model is shorter than the simulation time (1000 years), the ejection will repeat with a separation equal to the recurrence time, as listed in Table \ref{tab:sum}.
The advantage of the model grid is that (i) it covers a broad range of observed nova outburst characteristic, and (ii) it maps the physical properties of accreting white dwarfs, which can help model the nova population, particularly when a CV population is assumed or constrained.
When compared to other nova eruption models, the predicted recurrence time and ejecta mass from \citetalias{2005ApJ...623..398Y} model are in reasonable agreement with newer models \citep[e.g.,][]{2013ApJ...777..136W, 2019MNRAS.490.1678C, 2020NatAs...4..886H, 2021MNRAS.501..201H, 2024MNRAS.527.4806V} utilizing state-of-the-art stellar evolution codes.

Besides the parameters mentioned above, metallicity can also substantially affect properties of the nova outbursts.
However, \citetalias{2005ApJ...623..398Y} adopted a fixed metallicity at the solar abundance and assumed chemical diffusion on bare C/O cores.
Although this parameter is beyond the scope of this work, we note that lowering the metallicity to $\lesssim 0.1~Z_\odot$ can increase both the ejected mass and the kinectic energy of the eruption by factors of a few \citep{2007ApJ...662L.103J,2009ApJ...692..324S}.

\subsection{Photoionization from the post-eruption phase}\label{sec:photoionization}
During nova eruption, most accreted material would be ejected and stable thermonuclear burning is expected to occur on the remaining envelope subsequently, manifested as a supersoft X-ray source \citep{2013ApJ...777..136W}.
The post-eruption, supersoft phase generally has a duration between days to years, and a effective temperature of $10^5-10^6~\rm K$ \citep{2011A&A...533A..52H,2014A&A...563A...2H}.
Thus the emission from the supersoft phase could ionize and heat the expanding ejecta when they are close enough to the central WD.
By modelling thermal evolution of the post-eruption stage and photoionization of the expanding ejecta, \citet{2015ApJ...803...76C} demonstrated that the ejecta would undergo an isothermal phase with a temperature of $(1-4)\times\rm 10^4~\rm K$.
The duration of the isothermal phase is given by
\begin{equation}	
	t_{\rm iso}\approx236~{\rm days}\Big(\frac{M_{\rm ej}}{10^{-5}~\rm M_\odot}\Big)^{1/2}\Big(\frac{v_{\rm ej}}{10^3~\rm km~s^{-1}}\Big)^{-3/2}, 
\end{equation}
where $M_{\rm ej}$ and $v_{\rm ej}$ are mass and velocity of the ejecta, assuming a constant temperature, $T_{\rm iso}$, of $2\times10^4~\rm K$ and effective photon energy from the post-eruption phase of 22 eV.

To model the early isothermal expansion in our simulations, we impose a temperature floor, $T_{\rm iso}$, on the ejecta during the time interval $t_{\rm iso}$. 
Additionally, we incorporate the non-equilibrium ionization network from \citet{2008A&A...488..429T} to account for the recombining plasma in the expanding ejecta that follows the isothermal phase.
This ionization network includes all ions for H and He, as well as the first five (three) ions for C, N, and O (Ne and Fe). 
This approach self-consistently includes optically-thin radiative cooling resulting from non-equilibrium ionization within the temperature range $10^3~\rm K\leq T \leq 10^5~\rm K$, which is sufficiently broad to cover the expanding ejecta with a typical temperature of $T\sim10^4~\rm K$.
In the current cooling module, for gas at $T\gtrsim10^5~\rm K$ free-free emission dominates, while emission lines from highly ionized metals are omitted.

\begin{figure*}
 	\includegraphics[width=\linewidth]{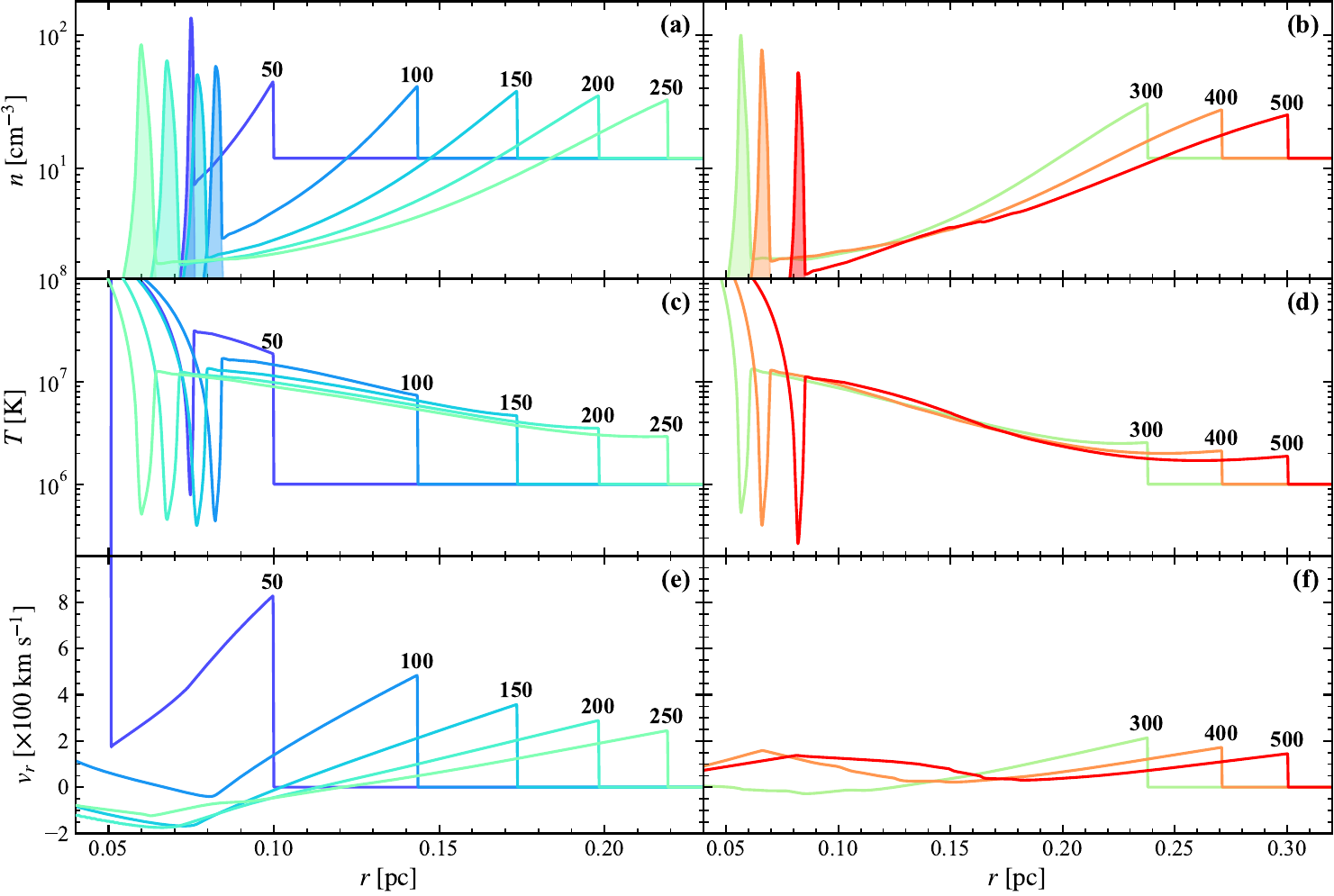}
	\caption{Snapshots of one model {\tt N06} for example. {\it Top}: number density distribution. Shaded region represents the density of ejecta component. {\it Middle}:  temperature distribution. {\it Bottom}:  radial velocity distribution, with positive value as outward motion. In all panels, the number indicates the time of snapshot in units of year. \label{fig:distribution}}
\end{figure*}

\subsection{Results} 
In Figure \ref{fig:distribution}, we show the evolutionary distributions of model {\tt N06} as an example, where the ejecta mass is $2.76\times10^{-4}~\rm M_\odot~yr^{-1}$ and the ejecta velocity is $2650~\rm km~s^{-1}$.
The distribution and evolution of the shocked ISM is similar to the classical Sedov-Taylor blast wave model.
Differences come from two assumptions: (i) the ISM pressure is neglected in Sedov-Taylor model while the hot and dense ISM can have considerable decelerating effects on the ejecta; (ii) the ejection is instantaneous in Sedov-Taylor model while the nova ejection is in the form of continuous and weakening wind.

As shown in Figure \ref{fig:distribution}, the swept ISM is heated to $\sim10^7~\rm K$ in the shock front before the ejecta is significantly decelerated, which dominates the heard X-ray emission.
As the shock becomes weaker, the subsequently shocked ISM has a temperature of $\sim10^{6}~\rm K$, which makes the remnant faint in hard X-rays.
During the collision between the ejecta and the ISM, the majority of ejecta is decelerated and pushed backward.
The backward ejecta compresses the later-generated ejecta and the consequent increase of pressure in the inner region would push forward the ejecta again (panel (e) and (f) in Figure \ref{fig:distribution}).
The dynamic balance between the pressure of the inner shocked ejecta and outer shocked ISM makes the ejecta bounded in a relatively thin shell, the position of which oscillates at a certain radial range.
The temperature of the ejecta is $\sim 2\times10^4$ K at the injection surface and remains nearly constant within $t_{\rm iso}$ in the first few years.
After the isothermal stage, the ejecta are then compressed and heated by the reverse shock, with temperature around $10^6$ K.
After that, the shell cools to several $10^5$ K due to expansion and radiative cooling, with the cooling time scale of 100--1000 years in the temperature range $5\times10^5$--$10^6$ K for typical densities of 10--100 $\rm cm^{-3}$.

\section{Multiwavelength synthesis}\label{sec:detect}
Galactic nova outbursts are observable across nearly the entire electromagnetic spectrum, from radio to gamma rays.
Dozens of resolved nebular remnants, with ages up to a century, have been identified, primarily in the optical band \citep{2008clno.book.....B}. 
Four nova remnants also exhibit resolved X-ray emission, which originates from the interaction between the ejected material and the circumstellar medium.
Among these, a notable example is nova GK Per (also known as Nova Persei 1901), whose century-old remnant features a unilateral X-ray shell dominated by non-equilibrium ionization thermal emission \citep{2005ApJ...627..933B}.

In the context of nova remnants in the GC, optical to soft X-ray ($\lesssim2~\rm keV$) emission is invisible due to prohibitive extinction, and not all novae are observed to have gamma-ray emission due to the limited flux sensitivity of present detectors \citep{2018A&A...609A.120F,2021ARA&A..59..391C}.
Consequently, emissions in the hard X-ray, radio, and infrared bands represent the most promising tracers for such remnants.

\subsection{Synthesis of hard X-ray emission}
At hard X-rays, about 4.5 Ms exposures toward the GC have been achieved by {\it Chandra} X-ray Observatory with the Advanced CCD Imaging Spectrometer (ACIS).
Within inner $500\arcsec$ of the GC, an unprecedented sensitivity for point source has been reached down to $0.9\times10^{-7}~{\rm photon~cm^{-2}~s^{-1}}$ at 2--8 keV, corresponding to $1.0\times10^{31}~{\rm erg~s^{-1}}$ at the distance of the GC.

To facilitate prediction for hard X-ray detectability of simulated nova remnants, we calculate the synthetic X-ray luminosity by
\begin{equation}
	L_{\rm X}=\int \Lambda_{\rm X} n_e n_{\rm H} 4\pi r^2dr,
\end{equation} 
where $\Lambda_{\rm X}$ is the X-ray emissivity at 2--8 keV in units of $\rm erg~cm^{3}~s^{-1}$, $n_e$ is electron density, $n_{\rm H}$ is hydrogen density.
As the time scale of temperature change in the shock front is much shorter than that of ionization equilibrium establishment, we employ the X-ray emissivity $\Lambda_{\rm X}$ of an optically-thin thermal plasma with non-equilibrium ionization (NEI), which consists of thermal bremsstrahlung emission and numerous ionic emission lines.
The emissivity is extracted from ATOMDB\footnote{\url{http://www.atomdb.org}}.
The NEI emissivity is a function of electron temperature $T_e$ and ionization timescale $n_e t$, which is the product of electron density and time elapsed since the shock passage.
We use passive tracer to calculate the ionization timescale on-the-fly in our simulations.
Specifically, one additional passive tracer $n_e t$ is added to the hydrodynamical equations in the form:
\begin{equation}
	\frac{\partial Q}{\partial t} + {\bm v}\cdot\nabla Q = S_Q,
\end{equation}
where $Q$ is the passive tracer and $S_Q$ is the source term which is $n_e$ for shocked gas and $0$ for unshocked gas. In each timestep of the simulation, whether the gas in each cell has been shocked is determined by the shock detector criteria according to \citet{2015MNRAS.446.3992S}. 

The NEI model is also parameterized by elemental abundances and initial ionization population.
The measurements on gas-phase elemental abundances in the GC are subject to large uncertainties due to the intense obscuration of valid abundance tracer in soft X-ray to optical band and the unconventional abundances of light elements.
We set the metallicity based on measurements of the dominant metal-rich stellar population in the NSC \citep{2023ApJ...944...79C}, assuming that shares a similar metallicity with this population.
However, in the inner parsec of the NSC, the ISM primarily originates from winds of Wolf-Rayet stars with totally different abundances, which are depleted with hydrogen and enriched with carbon and nitrogen.
\citet{2023MNRAS.522..635H} measured the abundances of gas-phase heavy elements in this region utilizing X-ray spectral analysis and found a sub-solar metallicity.
Moreover, the enrichment of specific elements due to supernovae, young stars, and giant stars would complicate the composition and augment uncertainty.
Considering the aforementioned, although the ISM elemental abundances of the NSC could be inhomogeneous and deviate from solar abundance ratio, we adopt a twice-solar metallicity and a solar abundance ratio ($\rm [M/Fe]=0$).
The treatment of metallicity could cause up to a threefold difference on the synthetic X-ray luminosity in 2--8 keV, with the maximum for supersolar metallicity and the minimum for hydrogen-depleted subsolar metallicity.
As for the initial ionization population, which is the ionization states of the gas before shock heating, it is assumed to stay in collisional ionization equilibrium at the temperature of the ISM as set in Section \ref{subsec:init}.

\begin{figure}
	\includegraphics[width=\linewidth]{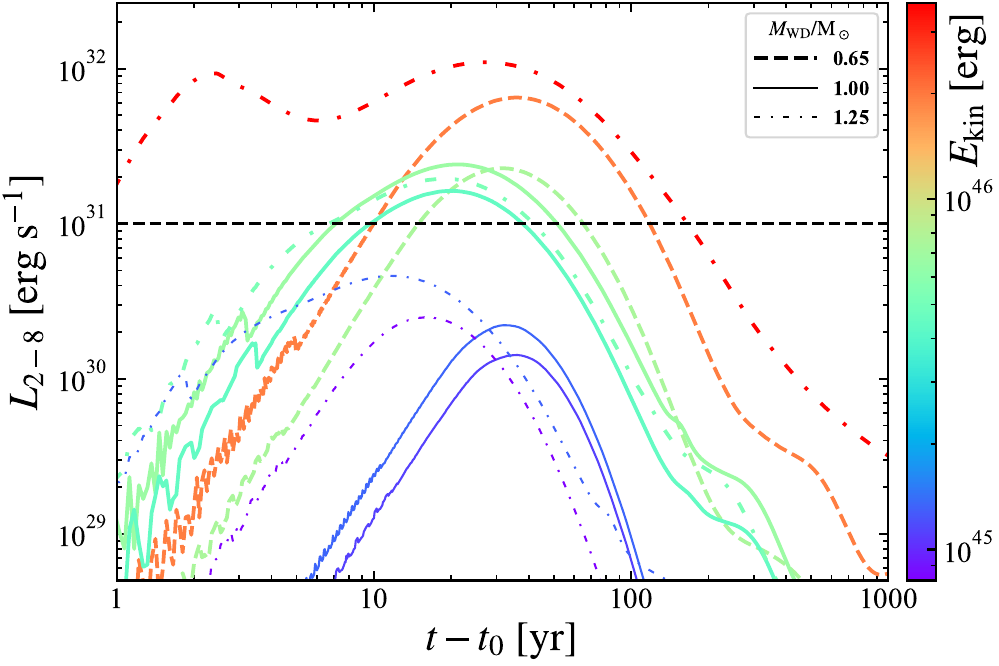}
	\caption{Synthetic X-ray light curve of simulated remnants with peaked 2--8 keV luminosities above $10^{30}~\rm erg~s^{-1}$. The time is relative to the ejection at the inner boundary. Dashed, solid, and dash-dotted lines indicate WD mass of 0.65, 1.00, and 1.25 solar masses, respectively. Colors of the curves represent kinetic energy of the ejecta. The horizontal, dashed line is the point-source detection limit of {\it Chandra} observations of the GC, $L_{\rm det}\approx10^{31}~\rm erg~s^{-1}$ \citep{2018ApJS..235...26Z}. \label{fig:lc}}
\end{figure}

If assume electron-ion temperature equilibrium $T_e=T$, 6 out of 79 nova remnants have X-ray luminosities that peak above the detection limit by the $\sim4.5~\rm Ms$ {\it Chandra} observations spanning 13.5 years.
These detectable remnants all have ejecta mass larger than $10^{-5}~\rm M_\odot$ among which the remnants with ejecta kinetic energy $\gtrsim5\times10^{45}~\rm erg$ can remain detectable for more than 100 years.
The synthetic X-ray light curves are generally below $10^{32}~\rm erg~s^{-1}$ (Figure \ref{fig:lc}), aligning with the observed X-ray luminosities of nova remnants, which range from $\sim10^{29}$ to $\sim10^{32}~\rm erg~s^{-1}$ \citep[e.g.,][]{2003ApJ...594..428M,2005ApJ...627..933B,2010MNRAS.404L..26B}.

\begin{figure}
	\includegraphics[width=0.95\linewidth]{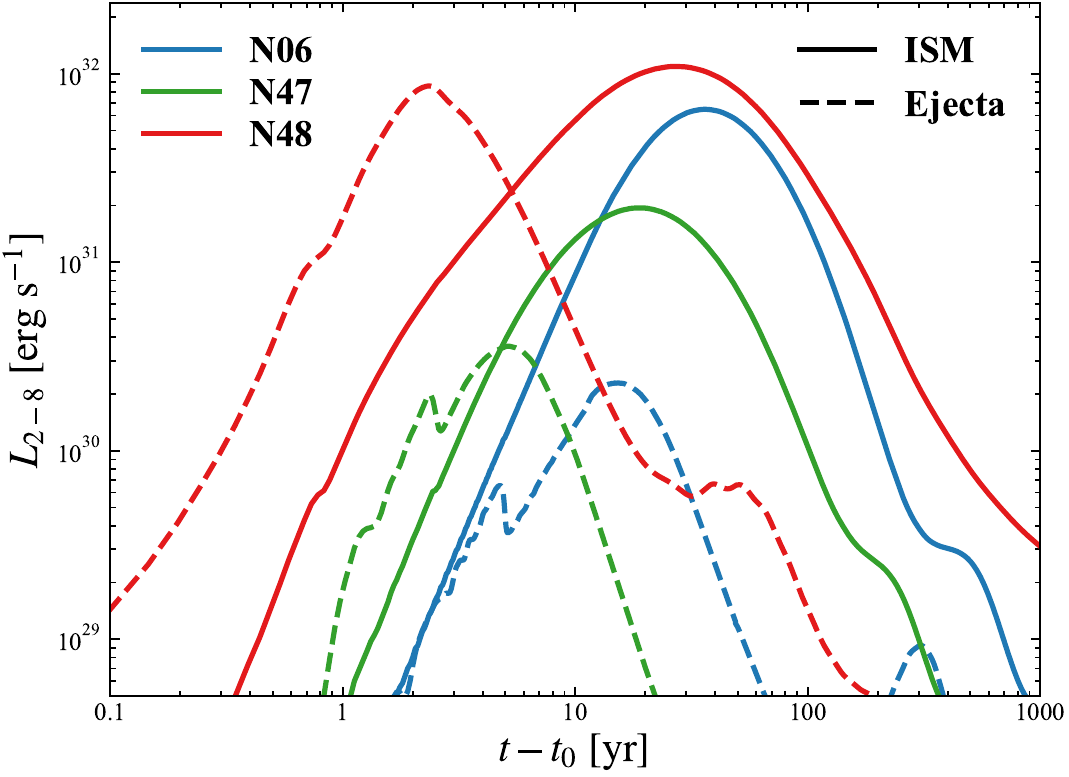}
	\caption{Synthetic X-ray light curves of the shocked ISM (solid curves) and ejecta (dashed curves) for three X-ray detectable models N06 (blue), N47 (green), and N48 (red). \label{fig:xray_component}}	
\end{figure}

Figure \ref{fig:xray_component} shows the contribution to hard X-ray from the shocked ISM and ejecta.
The long-term light curve is dominated by the shocked ISM, while the X-ray emission from the shocked ejecta evolves more rapidly and would be important for $t<10~\rm yr$.
Notably, for model N48 with the fastest ejecta, the X-ray maximum from the shocked ejecta is comparable to that of the shocked ISM, which is manifested by the two distinct peaks at $\sim10^{32}~\rm erg~s^{-1}$ in the collective light curve shown in Figure \ref{fig:lc}.

Observed hard X-ray emission in novae typically has a peak luminosity of $10^{33}-10^{34}~\rm erg~s^{-1}$ and remains detectable for weeks to months generally 1-2 months after the nova burst \citep{2008ApJ...677.1248M, 2021ApJ...910..134G}.
This early, rapid evolution in X-ray light curve is not reproduced in our simulated nova remnants because we do not include the mechanisms to generate the early X-ray emission, such as internal shocks in the ejecta, interaction between ejecta and dense circumstellar material, or enhanced accretion onto the white dwarf.
Additionally, searching for novae by X-ray transients is beyond the scope of this work since we focus on the faint but long-existing remnants that could be potentially detected by {\it Chandra}.
The fast X-ray evolution would be possibly missed by {\it Chandra} observations on the GC with an average cadence of 3 months.
Moreover, the {\it Swift} quasi-daily monitoring of the GC with a 2--10 keV sensitivity of $\simeq10^{34}~\rm erg~s^{-1}$ \citep{2015JHEAp...7..137D}, which is much shallower than the deep {\it Chandra} exposures but has much higher cadence, would be a rather better probe to capture the early, bright X-ray emission from novae.

\subsection{Synthesis of radio emission} \label{sec:radio}
Radio emission is often observed in nova bursts and their remnants, mostly arising due to thermal free-free emission from the warm, ionized ejecta shell with typical temperature of $10^4$--$10^5$ K.
Signatures of non-thermal radio emission, possibly related to shocks, also exhibit in a subset of novae.
However, non-thermal radio emission is found to only contribute in early times and the duration is much shorter than that of thermal emission \citep{2021ApJS..257...49C}.
The radio emission at 5 GHz is observed to peak at 0.1--100 mJy and decay over months to years \citep{2021ApJS..257...49C}.
Thus the radio emission from the nova remnants in the GC region maybe revealed by the recent deep radio observations that have achieved a sensitivity of a few $\mu\rm Jy~beam^{-1}$ utilizing the Jansky Very Large Array (JVLA) \citep{2020ApJ...905..173Z}.
Moreover, a population of compact radio sources with a size of $<1^{\prime\prime}$ have been found in the GC region, namely Galactic Center Compact Radio (GCCR) sources, some of which manifest as variables or transients \citep{2020ApJ...905..173Z,2022ApJ...927L...6Z}. 
We calculate the self-absorbed free-free emission from the simulated remnants at 5.5 ({\it C} band) GHz with existing JVLA observations.
We adopt the free-free emissivity $j_{\nu}$ and the attenuation coefficient $\kappa_{\nu}$ due to free-free absorption according to equation (10.1) and (10.13) in \citet{2011piim.book.....D}.

\begin{figure}
    \includegraphics[width=\linewidth]{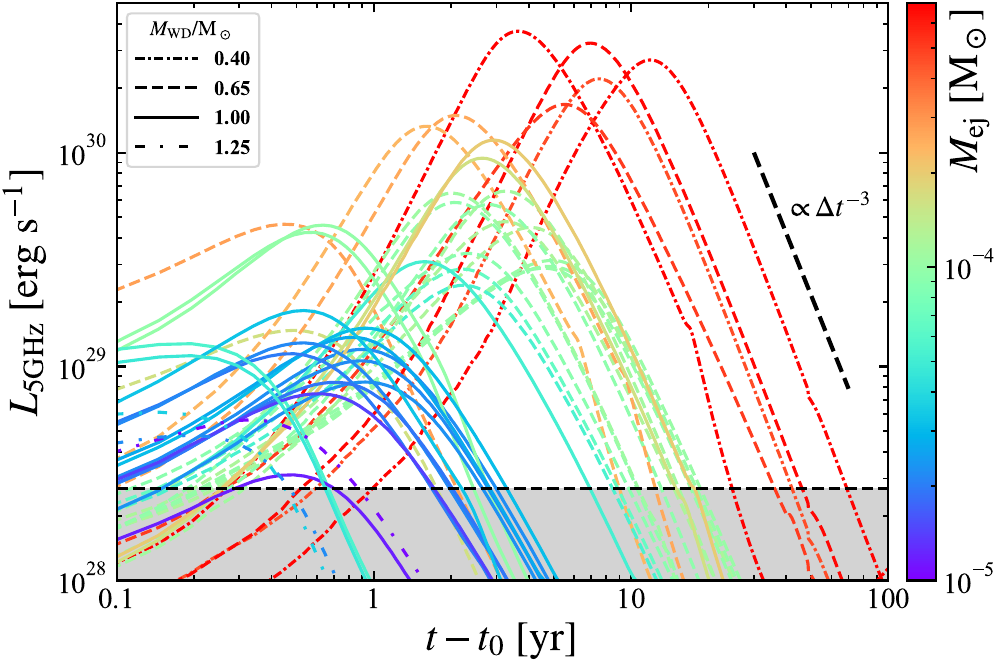}
    \caption{Synthetic radio light curve at 5 GHz ({\it C} band) of detectable simulated novae. The radio luminosity is defined as $\nu L_\nu$ at $\nu=5~\rm GHz$. The line styles indicate WD mass and colors of the curves represent ejecta mass. The horizontal, dashed line is the detection limit of JVLA observations of the GC, which is a $10\sigma$ detection limit of 70 $\mu\rm Jy$ \citep{2020ApJ...905..173Z}. The grey shaded region represents undetectable stages. \label{fig:radio_lc}}
\end{figure}

\begin{figure}
    \includegraphics[width=\linewidth]{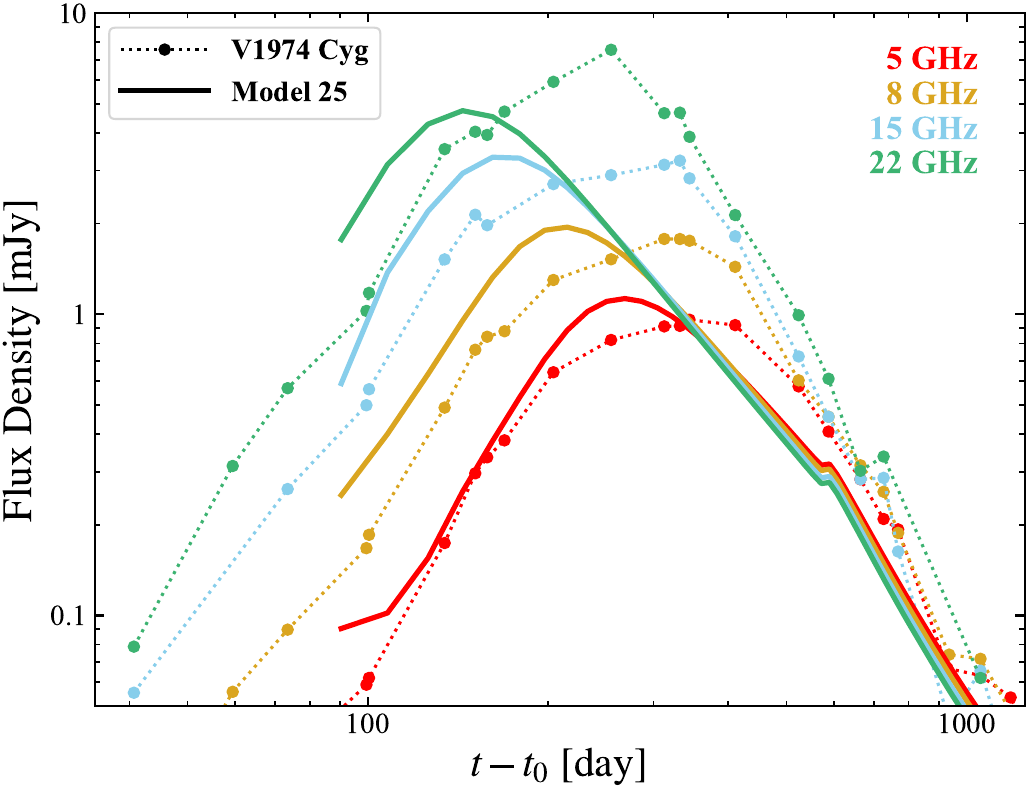}
    \caption{Comparison of radio light curve between simulated and observed novae with the distance of the GC assumed. The simulated nova (solid lines) is model {\tt N25} (see Table \ref{tab:sum}) in this work. Nova V1974 Cyg (points and dotted lines) is selected for comparison because of its complete light curves and thermal origin for radio emission, which is adopted from \citet{2021ApJS..257...49C}. Here $t_0$ is the time of ejection at the WD surface for simulated nova and the time of nova discovery for V1974 Cyg, respectively. The line colors indicate the frequencies. \label{fig:radio_compare}}
\end{figure}

In Figure \ref{fig:radio_lc}, we show the synthetic radio light curves at 5 GHz. 
The peak luminosities and duration of detection window are positively correlated with ejecta masses, implying the radio emission is dominated by the photoionized, warm ejecta.
The light curves of different models exhibit similar trend: fast rise as optical depth declines and slow decay due to ejecta expansion.
The decaying phase is characteristic of a power law with an index of $-3$, implying the optically thin decay.
More than half of nova models (44/79) are detectable for JVLA 5.5 GHz observations with detection window ranging from months to several ten years.
To evaluate the feasibility of the radio synthesis, we compare radio light curves of nova V1974 Cyg with a radio analogue, model {\tt N25}, in our simulations. 
As shown in Figure \ref{fig:radio_compare}, the turnover in the light curves and the frequency-dependent behavior are well reproduced, whereas the discrepancies in the peak time and the slope of decaying phase can be attributed to the assumed mass and velocity distribution of the ejecta.

\begin{figure}
    \includegraphics[width=\linewidth]{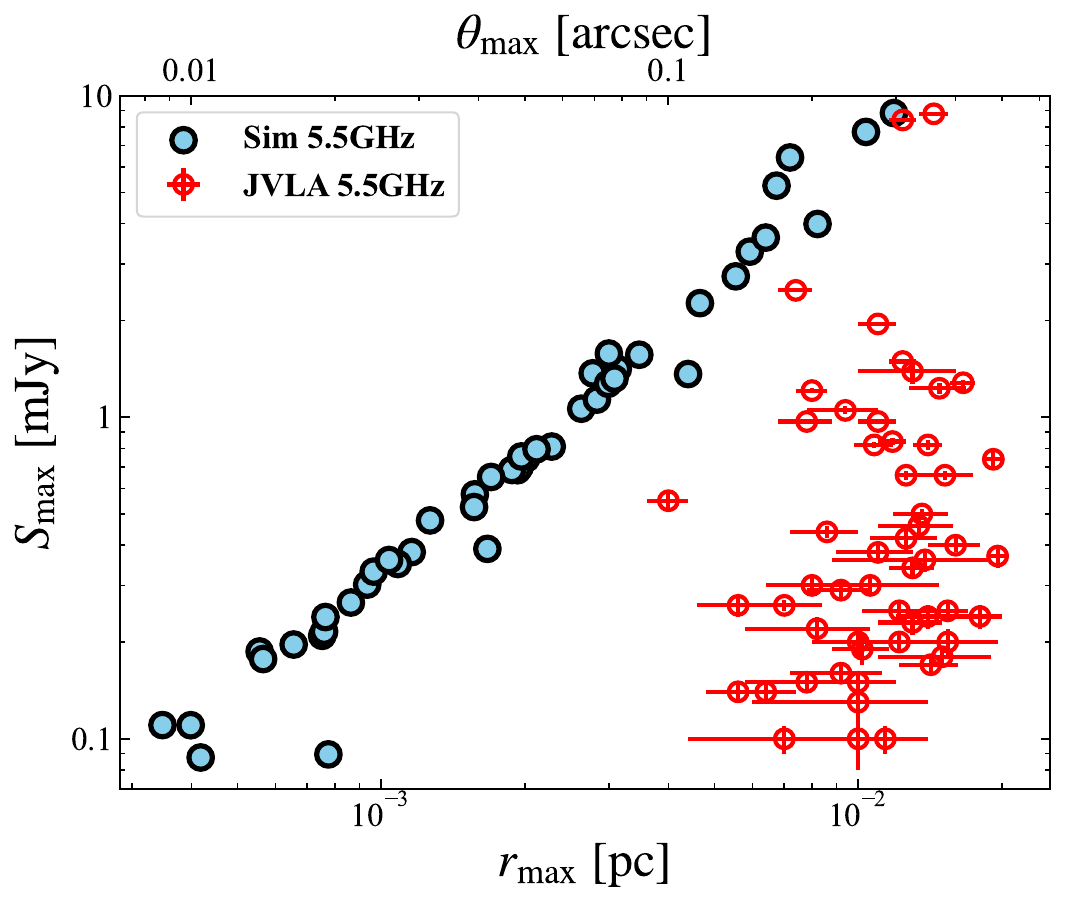}
    \caption{Flux density versus size of detectable remnants in the simulations and observed compact radio sources in the Galactic center. The filled circles represent the simulated remnants and the unfilled symbols with error bar represent JVLA measurements of the compact radio sources in the Galactic center \citep{2020ApJ...905..173Z}. For the simulated remnants, the flux density is for the radio maximum and the physical size is the maximal radius when the remnant is detectable.\label{fig:radio_size}}
\end{figure}

In Figure \ref{fig:radio_size}, we present the synthetic flux density and physical size of detectable nova remnants.
The majority of the simulated remnants have flux densities ranging from 0.1 to 10 mJy, comparable to those of the GCCR sources.
However, their physical sizes are essentially several times smaller than both the compact radio sources and the resolution limit of JVLA observations.
As a result, while nova remnants in the GC could be detected through deep JVLA observations, they would remain unresolved.
The apparent difference between the simulated remnants and the GCCR sources in Figure \ref{fig:radio_size} indicates that the majority of the GCCR sources are likely not associated with nova remnants (see Section \ref{sec:mc} and \ref{subsec:candidate} for further discussions).

\subsection{Synthesis of Pa$\alpha$ emission} \label{sec:Pa}
In infrared band, most known nova shells are discovered by emission lines from the warm ionized gas.
Hydrogen Pa$\alpha$ line at 1.87 $\rm \mu m$ are detected and spatially resolved in three nova shells, which are V1974 Cyg, QU Vul, and QV Vul \citep{2002AJ....124.2888K}.
These three remnants have Pa$\alpha$ line luminosities of $10^{31}-10^{32}~\rm erg~s^{-1}$ and spatial size of $0.01-0.03$ pc, assuming the distances\footnote{The distances are $1.61_{-0.12}^{+0.16}$ kpc for V1974 Cyg, $1.7_{-0.7}^{+1.2}$ kpc for QU Vul, and $3.6_{-1.1}^{+2.2}$ kpc for QV Vul, respectively.} estimated by \citet{2021AJ....161..147B} utilizing Gaia EDR3 data \citep{2016A&A...595A...1G,2021A&A...649A...1G}. 
In the GC region, {\it Hubble space telescope} (HST) narrow band observations have mapped Pa$\alpha$ emission with an angular resolution of $\sim0.2$ arcsec (corresponding to 0.008 pc at the GC distance) \citep{2010MNRAS.402..895W,2011MNRAS.417..114D} and would be capable to resolve the aforementioned nova shells if they existed in the GC.
Notably, the HST observations discovered several compact Pa$\alpha$ nebulae with angular scales ranging from sub-arcsecond to a few arcseconds, comparable to those observed for nova remnants. 
Examples include the ring-shaped nebula G-0.133-0.044 and G-0.240-0.082 (see Figure 5 in \citet{2010MNRAS.402..895W}).
This highlights the potential of HST Pa$\alpha$ mapping of the GC as a probe for detecting nova remnants through recombination lines.

To facilitate the comparison with observations and estimate detectability, we calculate the synthetic Pa$\alpha$ line emission of simulated novae.
We adopt the case B Pa$\alpha$ line emissivity in the form of,
\begin{eqnarray}
	4\pi j_{\rm Pa\alpha}&=&4.21\times10^{-26}\times \nonumber \\
	                 && T_4^{-1.122-0.039\ln T_4}n({\rm H^+}) n_e~\rm erg~s^{-1}~cm^3,
\end{eqnarray}
where $T_4$ is the electron temperature in unit of $10^4~\rm K$, $n({\rm H^+})$ and $n_e$ are number density of ionized hydrogen and electron.
The formula represents the best-fit result for the recombination rate from \citet{1995MNRAS.272...41S}, using fitting function from \citet{2011piim.book.....D}.

\begin{figure}
	\centering
	\includegraphics[width=\linewidth]{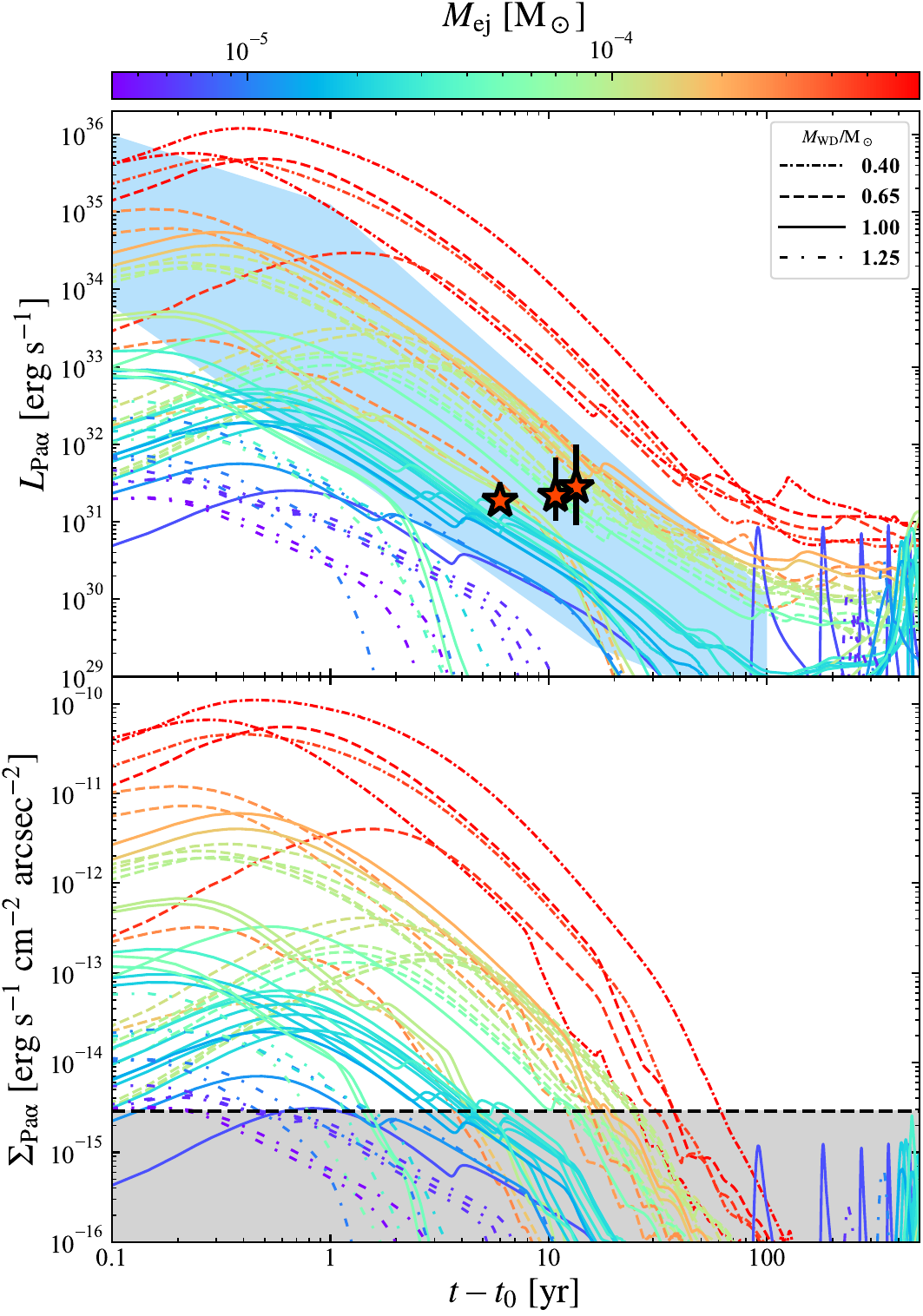}
    \caption{{\it Top}: Synthetic Pa$\alpha$ light curve of detectable simulated novae. The line styles and colors are as in Figure \ref{fig:radio_lc}. The blue shaded region outlines the $1\sigma$ uncertainty range for historical novae, based on H$\alpha$ light curve fits presented by \citet{2020A&A...641A.122T}. The stars, from left to right, represent V1974 Cyg, QV Vul, and QU Vul, with error bars indicating the $1\sigma$ uncertainties associated with their distances. {\it Bottom}: Extinction-inclusive Pa$\alpha$ surface brightness. The horizontal, dashed line indicates the $3\sigma$ sensitivity of the narrowband Pa$\alpha$ survey of the GC, at $2.9\times10^{-15}~\rm erg~s^{-1}~cm^2$ \citep{2010MNRAS.402..895W}. The grey shaded region marks undetectable stages. \label{fig:Pa_lc}}
\end{figure}

The case B recombination assumption is essentially reasonable for observed novae, as supported by line ratios from the hydrogen Brackett series in near-infrared band \citep[e.g.,][]{2015AJ....149..136R, 2015MNRAS.447..806D}, except during the very early evolutionary stages when the density is extremely high and/or the optical depth is large \citep{2000ApJ...541..791L}.
In our simulations, the electron density ($n_e<10^8~\rm cm^{-3}$) is far below the critical value for hydrogen Pa$\alpha$ line, $n_{\rm crit}\approx10^{11}~\rm cm^{-3}$, above which collisional deexcitation would overwhelm radiative transition.
As concerns the optical depth, we derive the maximal optical depth for Pa$\alpha$ line center using 
\begin{equation}
	\tau=\int_{r_{\rm in}}^{r_{\rm out}} n_e n_i \Omega dr,
\end{equation}
where $\Omega\approx3\times10^{-33}~\rm cm^{5}$ is the opacity factor under $T_e=10^4~\rm K$ and $n_e<10^8~\rm cm^{-3}$ according to \citet{1995MNRAS.272...41S}.
The optical depth derived for all snapshots of our simulations is smaller than 0.01, which indicates an optically thin condition for Pa$\alpha$ line.
Therefore, the case B assumption is valid in terms of the density and optical depth, which is a natural result as we do not include the early, optically thick stage due to the limitation of the resolution.
Furthermore, to consider the extinction toward the GC, we assume the average extinction $A_{\rm 1.87\mu m}=3.3$ in the central 5 pc.
This foreground extinction value is estimated by the 1.87$\mu$m-to-1.90$\mu$m flux ratio of point sources from \citet{2011MNRAS.417..114D}.
Finally, we derive the total intrinsic luminosity and the extinction-inclusive surface brightness of the simulated remnants (see Figure \ref{fig:Pa_lc}), with the surface brightness projected to match the 0.1 arcsec pixel size of HST/NICMOS.

As shown in Figure \ref{fig:Pa_lc}, the synthetic Pa$\alpha$ light curves have similar behavior to the radio light curves because they are both dominated by the warm ejecta.
However, the rising of Pa$\alpha$ emission is flatter than that of radio emission since self-absorption is negligible for Pa$\alpha$ line.
To compare the simulated light curve with observations, we utilize H$\alpha$ light curves of historical novae compiled by \citet{2020A&A...641A.122T}, considering the extensive H$\alpha$ monitoring of known novae and the tentative validity of case B assumption.
\citet{2020A&A...641A.122T} provide reasonable fits for the H$\alpha$ light curves of novae, grouped into four speed classes. 
These light curves exhibit a primary power-law decline with a slope ranging from $-3$ to $-2$, which are preceded and/or followed by relatively constant stages in some cases. 
We convert the observed H$\alpha$ luminosities to Pa$\alpha$ using the case B emissivity ratio, $j_{\rm Pa\alpha}/j_{\rm H\alpha} = 0.1$, at a temperature of $2\times10^4~\rm K$ \citep{1995MNRAS.272...41S}.
The luminosities and decline slopes of our simulated light curves are generally consistent with the $1\sigma$ uncertainty range of the fitted light curves for historical novae, depicted by the blue shaded region in Figure \ref{fig:Pa_lc}.
The Pa$\alpha$ luminosities and ages of resolved remnants, i.e., V1974 Cyg, QV Vul, and QU Vul, are also in line with our models with ejecta masses of $10^{-5}-10^{-4}~\rm M_\odot$.
Therefore, our simulations can reproduce the general behavior of Pa$\alpha$ emission of novae and would be plausible for further estimation of detectability.

As for the detectability, the majority of the simulated novae at their Pa$\alpha$ maximum (51/79) can be detected by the HST Pa$\alpha$ survey (Figure \ref{fig:Pa_lc}).
The duration of the detectable window spans from months to 60 years, with the longest for novae with massive and extremely slow ejecta.
The minimal ejecta mass for Pa$\alpha$ detection is $\sim3\times10^{-6}~\rm M_\odot$.
These results are also similar to the predictions for radio band in Section \ref{sec:radio}.

\begin{figure}
    \centering
	\includegraphics[width=\linewidth]{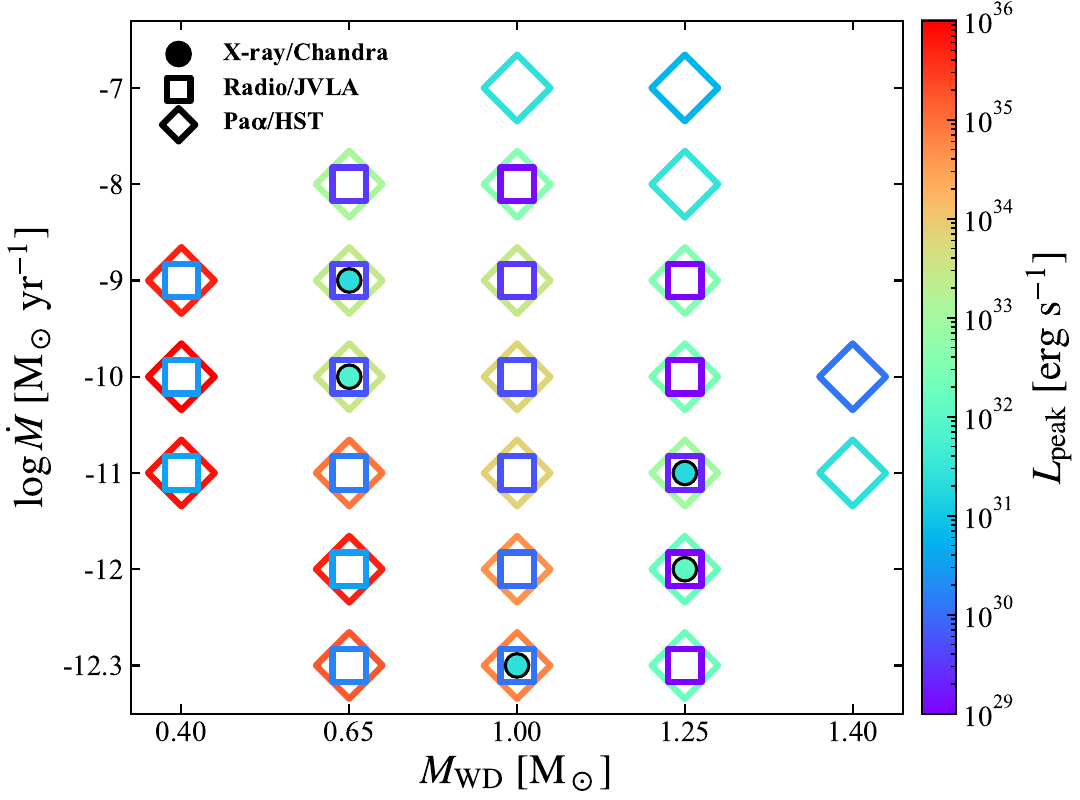}
    \caption{Summary of multiwavelength detectability as a function of mass transfer rate and masse of the WD. For different white dwarf temperatures, the model with the largest peak luminosity is taken. Circles, squares, and diamonds represent detectability for X-rays, radio band, and Pa$\alpha$ line, respectively. Colors indicate the luminosities at maximum.\label{fig:synthesis_sum}}
\end{figure}

To summarize, Figure \ref{fig:synthesis_sum} shows the multiwavelength detectability for our simulations. 
The X-ray detectable models only occupy the minority of all models, typically with ejecta velocity $v_{\rm ej}>2000~\rm km~s^{-1}$ and ejecta energy $E_{\rm ej}>4\times10^{45}~\rm erg$.
As for radio and Pa$\alpha$, detectable models cover the majority of the parameter space, and the peak luminosities increase with decreasing WD mass and decreasing mass transfer rate.
The parameters and multiwavelength luminosities of our simulations are detailed in Table \ref{tab:sum}.

\begin{table*}
\centering
\begin{threeparttable}
\caption{Summary of nova ejection properties and multiwavelength predictions\label{tab:sum}}
	{\footnotesize
	\centering
	\begin{tabular*}{\textwidth}{ccclcccc>{\centering\arraybackslash}p{1.0cm}c>{\centering\arraybackslash}p{0.6cm}c>{\centering\arraybackslash}p{0.6cm}}	
	\hline
	Model & $M_{\rm WD}$ & $T_{\rm WD}$ & $\log\dot{M}$ & $M_{\rm ej}$ & $P_{\rm rec}$ & $v_{\rm ej}$ & $L^{\rm peak}_{\rm X}$ & $t^{\rm det}_{\rm X}$ & $L^{\rm peak}_{\rm R}$ & $t^{\rm det}_{\rm R}$ & $L^{\rm peak}_{\rm Pa\alpha}$ & $t^{\rm det}_{\rm Pa\alpha}$ \\
	     & $(\rm M_\odot)$ & ($10^6~\rm K$) & $(\rm M_\odot~yr^{-1})$ & $(\rm M_\odot)$ & $\rm (yr)$ & $(\rm km~s^{-1})$ & $(\rm erg~s^{-1})$ & (yr) & $(\rm erg~s^{-1})$ & (yr) & $(\rm erg~s^{-1})$ & (yr) \\
	(1) & (2) & (3) & (4) & (5) & (6) & (7) & (8) & (9) & (10) & (11) & (12) & (13)\\
	 \hline
N01 & 0.40 & 10 & -9    & 4.6E-04 & 4.12E+05 & 149 & 8.8E+23 & $\cdots$ & 2.2E+30 & 40.08    & 5.2E+35 & 37.62    \\
N02 & 0.40 & 10 & -10   & 6.8E-04 & 5.62E+06 & 97 & 4.3E+23 & $\cdots$ & 2.7E+30 & 65.38    & 1.3E+36 & 61.04    \\
N03 & 0.40 & 10 & -11   & 7.0E-04 & 5.87E+07 & 462 & 1.9E+28 & $\cdots$ & 3.7E+30 & 23.65    & 7.0E+35 & 43.85    \\
N04 & 0.65 & 10 & -8    & 1.0E-04 & 1.01E+04 & 122 & 8.4E+22 & $\cdots$ & 3.1E+29 & 17.77    & 1.2E+33 & 25.60    \\
N05 & 0.65 & 10 & -9    & 1.6E-04 & 1.61E+05 & 2150 & 2.3E+31 & 50.13    & 1.6E+29 & 1.61     & 4.2E+32 & 4.15     \\
N06 & 0.65 & 10 & -10   & 2.8E-04 & 2.55E+06 & 2650 & 6.5E+31 & 107.82   & 5.8E+29 & 1.95     & 2.6E+33 & 3.67     \\
N07 & 0.65 & 10 & -11   & 2.4E-04 & 2.58E+07 & 623 & 1.1E+29 & $\cdots$ & 1.4E+30 & 8.60     & 8.3E+34 & 30.65    \\
N08 & 0.65 & 10 & -12   & 6.7E-04 & 3.94E+08 & 216 & 1.5E+25 & $\cdots$ & 3.3E+30 & 43.96    & 5.1E+35 & 37.38    \\
N09 & 0.65 & 10 & -12.3 & 5.0E-04 & 1.08E+09 & 230 & 1.0E+25 & $\cdots$ & 1.7E+30 & 34.65    & 3.0E+34 & 31.03    \\
N10 & 0.65 & 30 & -8    & 1.0E-04 & 1.02E+04 & 125 & 8.8E+22 & $\cdots$ & 2.9E+29 & 17.15    & 1.1E+33 & 24.72    \\
N11 & 0.65 & 30 & -9    & 1.2E-04 & 1.11E+05 & 156 & 2.2E+23 & $\cdots$ & 4.5E+29 & 16.52    & 2.7E+33 & 23.70    \\
N12 & 0.65 & 30 & -10   & 1.2E-04 & 9.55E+05 & 195 & 2.1E+24 & $\cdots$ & 4.8E+29 & 13.93    & 3.2E+33 & 19.83    \\
N13 & 0.65 & 30 & -11   & 9.5E-05 & 5.96E+06 & 189 & 6.3E+23 & $\cdots$ & 5.8E+29 & 12.32    & 2.3E+34 & 17.64    \\
N14 & 0.65 & 30 & -12   & 9.6E-05 & 4.12E+07 & 268 & 2.7E+26 & $\cdots$ & 6.0E+29 & 9.87     & 2.4E+34 & 15.59    \\
N15 & 0.65 & 30 & -12.3 & 1.1E-04 & 8.89E+07 & 297 & 3.3E+26 & $\cdots$ & 6.6E+29 & 10.07    & 2.8E+34 & 14.91    \\
N16 & 0.65 & 50 & -8    & 9.9E-05 & 1.06E+04 & 130 & 9.9E+22 & $\cdots$ & 2.9E+29 & 16.32    & 1.0E+33 & 23.61    \\
N17 & 0.65 & 50 & -9    & 9.2E-05 & 7.41E+04 & 240 & 9.7E+26 & $\cdots$ & 3.4E+29 & 9.29     & 1.6E+33 & 14.23    \\
N18 & 0.65 & 50 & -10   & 6.7E-05 & 5.22E+05 & 208 & 8.6E+25 & $\cdots$ & 2.9E+29 & 8.41     & 1.4E+33 & 12.65    \\
N19 & 0.65 & 50 & -11   & 5.4E-05 & 3.86E+06 & 179 & 2.6E+24 & $\cdots$ & 2.4E+29 & 8.02     & 1.1E+33 & 12.07    \\
N20 & 0.65 & 50 & -12   & 1.1E-04 & 4.54E+07 & 176 & 4.5E+23 & $\cdots$ & 6.7E+29 & 14.41    & 3.2E+34 & 20.04    \\
N21 & 0.65 & 50 & -12.3 & 2.6E-04 & 2.33E+08 & 487 & 1.3E+28 & $\cdots$ & 1.5E+30 & 11.97    & 1.6E+35 & 32.16    \\
N22 & 1.00 & 10 & -7    & 7.2E-06 & 8.96E+01 & 240 & 4.1E+26 & $\cdots$ & 1.8E+28 & $\cdots$ & 2.6E+31 & 8.29     \\
N23 & 1.00 & 10 & -8    & 2.2E-05 & 2.06E+03 & 271 & 8.4E+26 & $\cdots$ & 1.3E+29 & 2.79     & 3.9E+32 & 4.50     \\
N24 & 1.00 & 10 & -9    & 5.2E-05 & 4.66E+04 & 256 & 8.5E+26 & $\cdots$ & 3.2E+29 & 6.50     & 3.0E+33 & 13.39    \\
N25 & 1.00 & 10 & -10   & 9.7E-05 & 8.40E+05 & 1180 & 2.2E+30 & $\cdots$ & 5.4E+29 & 2.30     & 6.1E+33 & 4.30     \\
N26 & 1.00 & 10 & -11   & 1.0E-04 & 8.72E+06 & 1063 & 1.4E+30 & $\cdots$ & 5.6E+29 & 2.64     & 7.0E+33 & 4.74     \\
N27 & 1.00 & 10 & -12   & 1.6E-04 & 9.28E+07 & 292 & 4.1E+26 & $\cdots$ & 9.6E+29 & 14.03    & 4.1E+34 & 17.74    \\
N28 & 1.00 & 10 & -12.3 & 2.0E-04 & 2.08E+08 & 295 & 2.5E+26 & $\cdots$ & 1.2E+30 & 16.22    & 6.0E+34 & 18.86    \\
N29 & 1.00 & 30 & -7    & 5.7E-06 & 8.74E+01 & 237 & 3.5E+26 & $\cdots$ & 1.2E+28 & $\cdots$ & 1.7E+31 & $\cdots$ \\
N30 & 1.00 & 30 & -8    & 2.2E-05 & 2.03E+03 & 274 & 1.1E+27 & $\cdots$ & 1.1E+29 & 2.54     & 2.9E+32 & 4.20     \\
N31 & 1.00 & 30 & -9    & 3.1E-05 & 2.70E+04 & 324 & 1.5E+27 & $\cdots$ & 1.6E+29 & 3.03     & 5.4E+32 & 6.30     \\
N32 & 1.00 & 30 & -10   & 2.7E-05 & 2.10E+05 & 344 & 1.4E+27 & $\cdots$ & 1.5E+29 & 2.59     & 4.7E+32 & 5.72     \\
N33 & 1.00 & 30 & -11   & 2.0E-05 & 1.15E+06 & 446 & 4.7E+26 & $\cdots$ & 1.4E+29 & 1.52     & 1.2E+33 & 3.76     \\
N34 & 1.00 & 30 & -12   & 2.2E-05 & 9.72E+06 & 414 & 2.5E+26 & $\cdots$ & 1.5E+29 & 1.76     & 1.4E+33 & 4.35     \\
N35 & 1.00 & 30 & -12.3 & 5.5E-05 & 6.42E+07 & 3490 & 2.4E+31 & 44.22    & 3.4E+29 & 0.49     & 2.8E+33 & 1.37     \\
N36 & 1.00 & 50 & -7    & 5.0E-06 & 8.30E+01 & 239 & 4.1E+26 & $\cdots$ & 7.2E+27 & $\cdots$ & 1.0E+31 & $\cdots$ \\
N37 & 1.00 & 50 & -8    & 2.3E-05 & 2.27E+03 & 272 & 1.5E+27 & $\cdots$ & 8.8E+28 & 2.35     & 1.8E+32 & 3.91     \\
N38 & 1.00 & 50 & -9    & 1.9E-05 & 1.62E+04 & 328 & 1.3E+27 & $\cdots$ & 9.1E+28 & 1.76     & 2.0E+32 & 3.08     \\
N39 & 1.00 & 50 & -10   & 1.4E-05 & 1.09E+05 & 302 & 7.2E+26 & $\cdots$ & 8.3E+28 & 1.52     & 2.0E+32 & 2.64     \\
N40 & 1.00 & 50 & -11   & 1.2E-05 & 7.94E+05 & 471 & 1.3E+27 & $\cdots$ & 3.7E+28 & 0.44     & 6.0E+31 & 1.12     \\
N41 & 1.00 & 50 & -12   & 3.3E-05 & 2.07E+07 & 597 & 1.2E+28 & $\cdots$ & 2.2E+29 & 1.91     & 2.6E+33 & 7.72     \\
N42 & 1.00 & 50 & -12.3 & 4.7E-05 & 5.05E+07 & 3083 & 1.6E+31 & 28.26    & 2.8E+29 & 0.49     & 2.2E+33 & 1.47     \\
N43 & 1.25 & 10 & -7    & 1.6E-06 & 1.92E+01 & 346 & 1.2E+26 & $\cdots$ & 2.4E+27 & $\cdots$ & 3.5E+30 & $\cdots$ \\
N44 & 1.25 & 10 & -8    & 3.9E-06 & 3.67E+02 & 568 & 2.2E+27 & $\cdots$ & 1.8E+28 & $\cdots$ & 3.0E+31 & $\cdots$ \\
N45 & 1.25 & 10 & -9    & 1.1E-05 & 9.27E+03 & 413 & 2.0E+26 & $\cdots$ & 7.9E+28 & 0.68     & 3.5E+32 & 1.42     \\
N46 & 1.25 & 10 & -10   & 2.2E-05 & 1.91E+05 & 1940 & 2.5E+30 & $\cdots$ & 1.0E+29 & 0.34     & 2.5E+32 & 1.51     \\
N47 & 1.25 & 10 & -11   & 3.6E-05 & 2.97E+06 & 3857 & 2.0E+31 & 35.05    & 2.0E+29 & 0.29     & 8.8E+32 & 1.12     \\
N48 & 1.25 & 10 & -12   & 5.3E-05 & 3.22E+07 & 8267 & 1.1E+32 & 165.03   & 3.8E+28 & 0.15     & 1.6E+32 & 0.73     \\
N49 & 1.25 & 30 & -7    & 1.9E-06 & 1.96E+01 & 373 & 2.4E+26 & $\cdots$ & 3.2E+27 & $\cdots$ & 4.8E+30 & $\cdots$ \\
N50 & 1.25 & 30 & -8    & 4.2E-06 & 3.84E+02 & 531 & 4.6E+27 & $\cdots$ & 1.7E+28 & $\cdots$ & 2.6E+31 & $\cdots$ \\
N51 & 1.25 & 30 & -9    & 5.9E-06 & 5.22E+03 & 589 & 2.0E+27 & $\cdots$ & 2.9E+28 & $\cdots$ & 4.8E+31 & 0.39     \\
N52 & 1.25 & 30 & -10   & 5.6E-06 & 4.35E+04 & 678 & 5.6E+27 & $\cdots$ & 2.9E+28 & $\cdots$ & 5.1E+31 & 0.29     \\
N53 & 1.25 & 30 & -11   & 4.3E-06 & 2.34E+05 & 1264 & 1.1E+29 & $\cdots$ & 2.5E+28 & $\cdots$ & 3.9E+31 & 0.24     \\
N54 & 1.25 & 30 & -12   & 4.2E-06 & 2.10E+06 & 1696 & 3.1E+29 & $\cdots$ & 2.1E+28 & $\cdots$ & 3.3E+31 & 0.20     \\
N55 & 1.25 & 30 & -12.3 & 9.9E-06 & 1.16E+07 & 3668 & 4.6E+30 & $\cdots$ & 4.7E+28 & 0.05     & 1.7E+32 & 0.44     \\
N56 & 1.25 & 50 & -7    & 1.8E-06 & 1.96E+01 & 355 & 1.3E+26 & $\cdots$ & 4.1E+27 & $\cdots$ & 5.9E+30 & $\cdots$ \\
N57 & 1.25 & 50 & -8    & 4.0E-06 & 3.69E+02 & 533 & 6.4E+27 & $\cdots$ & 1.3E+28 & $\cdots$ & 1.9E+31 & $\cdots$ \\
N58 & 1.25 & 50 & -9    & 3.6E-06 & 3.18E+03 & 485 & 4.2E+26 & $\cdots$ & 1.6E+28 & $\cdots$ & 2.4E+31 & $\cdots$ \\
N59 & 1.25 & 50 & -10   & 2.8E-06 & 2.14E+04 & 542 & 4.3E+26 & $\cdots$ & 8.2E+27 & $\cdots$ & 1.2E+31 & $\cdots$ \\
N60 & 1.25 & 50 & -11   & 2.5E-06 & 1.62E+05 & 658 & 1.9E+27 & $\cdots$ & 6.4E+27 & $\cdots$ & 9.1E+30 & $\cdots$ \\
	\hline
	\end{tabular*}
	}
	\end{threeparttable}
\end{table*}

\begin{table*}
	\centering
	\begin{threeparttable}
	\contcaption{\label{tab:con}}
	{\footnotesize
	\centering
	\begin{tabular*}{\textwidth}{ccclcccc>{\centering\arraybackslash}p{1.0cm}c>{\centering\arraybackslash}p{0.6cm}c>{\centering\arraybackslash}p{0.6cm}}
	\hline
	Model & $M_{\rm WD}$ & $T_{\rm WD}$ & $\log\dot{M}$ & $M_{\rm ej}$ & $P_{\rm rec}$ & $v_{\rm ej}$ & $L^{\rm peak}_{\rm X}$ & $t^{\rm det}_{\rm X}$ & $L^{\rm peak}_{\rm R}$ & $t^{\rm det}_{\rm R}$ & $L^{\rm peak}_{\rm Pa\alpha}$ & $t^{\rm det}_{\rm Pa\alpha}$ \\
	     & $(\rm M_\odot)$ & ($10^6~\rm K$) & $(\rm M_\odot~yr^{-1})$ & $(\rm M_\odot)$ & $\rm (yr)$ & $(\rm km~s^{-1})$ & $(\rm erg~s^{-1})$ & (yr) & $(\rm erg~s^{-1})$ & (yr) & $(\rm erg~s^{-1})$ & (yr) \\
	(1) & (2) & (3) & (4) & (5) & (6) & (7) & (8) & (9) & (10) & (11) & (12) & (13)\\
	 \hline
N61 & 1.25 & 50 & -12   & 5.9E-06 & 4.43E+06 & 751 & 1.1E+28 & $\cdots$ & 4.7E+28 & 0.10     & 1.1E+32 & 0.44     \\
N62 & 1.25 & 50 & -12.3 & 1.0E-05 & 1.11E+07 & 1675 & 7.2E+29 & $\cdots$ & 7.5E+28 & 0.15     & 1.7E+32 & 0.63     \\
N63 & 1.40 & 10 & -7    & 5.3E-08 & 7.71E-01 & 681 & 1.2E+28 & $\cdots$ & 2.8E+25 & $\cdots$ & 4.4E+28 & $\cdots$ \\
N64 & 1.40 & 10 & -8    & 1.8E-07 & 1.64E+01 & 1160 & 1.3E+28 & $\cdots$ & 2.0E+26 & $\cdots$ & 3.1E+29 & $\cdots$ \\
N65 & 1.40 & 10 & -9    & 4.7E-07 & 4.12E+02 & 1020 & 8.3E+27 & $\cdots$ & 5.2E+26 & $\cdots$ & 9.4E+29 & $\cdots$ \\
N66 & 1.40 & 10 & -10   & 6.9E-07 & 5.90E+03 & 2020 & 7.9E+28 & $\cdots$ & 6.2E+26 & $\cdots$ & 1.2E+30 & $\cdots$ \\
N67 & 1.40 & 10 & -11   & 3.0E-06 & 2.59E+05 & 2817 & 8.4E+29 & $\cdots$ & 9.1E+27 & $\cdots$ & 2.5E+31 & 0.05     \\
N68 & 1.40 & 30 & -7    & 5.5E-08 & 7.94E-01 & 723 & 1.6E+28 & $\cdots$ & 3.5E+25 & $\cdots$ & 4.5E+28 & $\cdots$ \\
N69 & 1.40 & 30 & -8    & 2.0E-07 & 2.02E+01 & 840 & 5.1E+27 & $\cdots$ & 2.2E+26 & $\cdots$ & 3.7E+29 & $\cdots$ \\
N70 & 1.40 & 30 & -9    & 3.0E-07 & 2.64E+02 & 1800 & 2.5E+28 & $\cdots$ & 5.4E+26 & $\cdots$ & 8.3E+29 & $\cdots$ \\
N71 & 1.40 & 30 & -10   & 2.7E-07 & 2.11E+03 & 2120 & 3.6E+28 & $\cdots$ & 9.2E+25 & $\cdots$ & 1.4E+29 & $\cdots$ \\
N72 & 1.40 & 30 & -11   & 2.1E-07 & 1.28E+04 & 2738 & 4.9E+28 & $\cdots$ & 3.5E+25 & $\cdots$ & 4.6E+28 & $\cdots$ \\
N73 & 1.40 & 30 & -12   & 1.9E-07 & 9.30E+04 & 2312 & 3.0E+28 & $\cdots$ & 3.5E+25 & $\cdots$ & 4.6E+28 & $\cdots$ \\
N74 & 1.40 & 50 & -7    & 5.7E-08 & 8.09E-01 & 812 & 2.8E+28 & $\cdots$ & 3.9E+25 & $\cdots$ & 4.3E+28 & $\cdots$ \\
N75 & 1.40 & 50 & -8    & 2.0E-07 & 2.02E+01 & 833 & 5.0E+27 & $\cdots$ & 1.8E+26 & $\cdots$ & 2.8E+29 & $\cdots$ \\
N76 & 1.40 & 50 & -9    & 2.2E-07 & 1.90E+02 & 1580 & 1.2E+28 & $\cdots$ & 7.7E+25 & $\cdots$ & 1.3E+29 & $\cdots$ \\
N77 & 1.40 & 50 & -10   & 1.6E-07 & 1.21E+03 & 1870 & 1.5E+28 & $\cdots$ & 3.5E+25 & $\cdots$ & 4.8E+28 & $\cdots$ \\
N78 & 1.40 & 50 & -11   & 1.2E-07 & 6.83E+03 & 1900 & 1.2E+28 & $\cdots$ & 2.0E+25 & $\cdots$ & 2.6E+28 & $\cdots$ \\
N79 & 1.40 & 50 & -12   & 3.2E-07 & 2.44E+05 & 2570 & 6.4E+28 & $\cdots$ & 9.2E+25 & $\cdots$ & 1.4E+29 & $\cdots$ \\ 
	\hline
	\end{tabular*}
	}
	\begin{tablenotes}
	\item[] {\bf Notes.} (1) Model name. (2)--(4) Mass, core temperature, and logarithm of the mass transfer rate of the accreting white dwarf. (5)--(7) Ejecta mass, recurrence time,, and ejecta velocity of the nova, adopted from \protect\citetalias{2005ApJ...623..398Y}. (8)--(13) Maximum luminosities and length of detectable window for X-rays (2--8 keV), radio band (5 GHz) and Pa$\alpha$.
	\end{tablenotes}
	\end{threeparttable}
\end{table*}

\section{probability of detection} \label{sec:prob}
In Section \ref{sec:detect}, we have calculated multiwavelength luminosities and light curves for simulated nova remnants, parameterized by the mass, temperature, and mass transfer rate of white dwarfs.
To further estimate the realistic probability of detection, we have to consider the CV population in the NSC.

\subsection{CV population modelling} \label{sec:pop}

Observational determination of the CV population in the NSC is infeasible for the lack of optical information.
In terms of theoretical predictions, CV and nova population synthesis is also a complicated problem concerning binary formation and evolution \citep[e.g.,][]{2016MNRAS.458.2916C, 2021MNRAS.504.6117K}, especially when dynamical interaction in the NSC entangles the case.
Here, we simply model the CV population based on assumptions on three key parameters, the WD mass and mass transfer rate distribution, and the CV count.

For the WD mass distribution, we refer to the local CV sample whose properties are better constrained due to proximity and minimal extinction.
The local CVs $(d\lesssim300~\rm pc)$ have a mean mass of $\langle M_{\rm WD}\rangle\approx0.8~\rm M_\odot$ and the mass spectrum concentrates within $0.6-1.0~\rm M_\odot$ \citep{2022MNRAS.510.6110P}.
Moreover, the X-ray measurement on the mean WD mass in the Galactic bulge CVs, $0.81\pm0.07~\rm M_\odot$, is also well consistent with local CVs \citep{2018ApJ...853..182Y}.
We thus assume a simple normal distribution with mean WD mass of $0.8~\rm M_\odot$ and standard deviation of $0.15~\rm M_\odot$ to model the WD mass distribution, called ``fiducial distribution''.
However, the NSC is suggested to host a population of CVs with massive WDs ($M_{\rm WD}>1.0~\rm M_\odot$), which are rare in the field and imply a different WD mass distribution \citep{2019ApJ...882..164X}.
The different CV population is also supported by the steeper flux distribution of X-ray sources in the NSC than in the Galactic bulge \citep{2018ApJS..235...26Z}.
Therefore, we tentatively assume a ``NSC distribution'' with the same distribution as ``fiducial distribution'' but higher mean WD mass of $1.0~\rm M_\odot$.

For the WD transfer rate, we adopt a log-normal distribution with mean mass transfer rate of $\langle\dot{M}\rangle=5.0\times10^{-10}~\rm M_\odot~yr^{-1}$ and standard deviation of 0.5 dex, which is broadly consistent with the local sample \citep{2022MNRAS.510.6110P}.

For the number of CVs in the NSC, we start with the number of X-ray sources detected by {\it Chandra} deep observations on the inner 500\arcsec ($\sim20~\rm pc$) of the GC \citep{2018ApJS..235...26Z}, which is around 3600.
The vast majority of these X-ray sources are believed to be CVs, while a portion of CVs are missing from detection due to incompleteness below the sensitivity limit of $10^{31}~\rm erg~s^{-1}$.
The undetected CVs are mainly faint non-magnetic CVs as most magnetic CVs tend to be detectable due to relatively high X-ray luminosities ($L_{\rm X}\sim10^{31}-10^{34}~\rm erg~s^{-1}$) and hard thermal bremsstrahlung spectra ($kT\gtrsim10~\rm keV$) \citep{2016ApJ...818..136X, 2024MNRAS.527.7173B}.
If roughly assuming that all magnetic CVs and part of non-magnetic CVs with fraction $f_{\rm det}$ are detected and that magnetic CVs make up 30\% of all CVs, the number of all CVs would be 7600 (12000) for $f_{\rm det}=0.25~(0)$.
Therefore, we take the number of all CVs in the NSC to be $10^4$, corresponding to a CV abundance\footnote{$M_{\rm NSC}=2.5\times10^{7}~\rm M_\odot$ is the mass of the NSC \citep{2014A&A...566A..47S}.} $\nu_{\rm CV}=N_{\rm CV}/M_{\rm NSC}=4.0\times10^{-4}~\rm M_{\rm \odot}^{-1}$.
This abundance is in line with and about two times that of the Galactic field of $1.7\times10^{-4}~\rm M_{\odot}^{-1}$ \citep{2023AJ....165..163C} \footnote{The local CV spatial density from \citet{2023AJ....165..163C} is translated to the CV abundance by assuming a local stellar density $0.04~\rm M_\odot~pc^{-3}$ \citep{2017MNRAS.470.1360B}.}.  

Compared with our assumption here, recent Monte Carlo simulations of realistic dense star clusters predict similar or higher CV mass density.
For example, based on Monte Carlo simulations for 288 globular cluster models, \citet{2019MNRAS.483..315B} show correlation between CV number and cluster mass with $\nu_{\rm CV}=2.11\times10^{-4}-2.66\times10^{-3}~\rm M_{\odot}^{-1}$ for different setup of initial binary population and common envelop phase efficiency.
Moreover, \citet{2022ApJ...931...84Y} predict 334 CVs in simulated 47 Tuc analogue, giving $\nu_{\rm CV}=3.3\times10^{-4}~\rm M_{\odot}^{-1}$.
Although these simulations are performed for globular clusters and sensitive to specific parameters, the results indicate that the CV number in the NSC could be several times higher than our estimation.

Furthermore, although there is a lack of nova detection in nuclear star clusters, the observed nova rates in globular clusters in nearby galaxies are suggested to be up to one order of magnitude higher than that in the galactic field \citep{2013A&A...549A.120H,2015ApJ...811...34C}.
In spite of large uncertainties on these estimations bases on scanty sample of detected novae in globular clusters, this indicates tentatively higher nova rate and/or CV abundance in compact stellar systems such as the NSC.
Considering the results from simulations on clusters and monitoring on novae, our assumption here may therefore be conservative.
A sophisticated modelling of the CV population in the NSC is beyond the scope of this research and is rather difficult solely based on present data, considering possible difference between the CV population in the NSC and the Galactic disk/bulge.

\subsection{Monte Carlo simulation on detection probability} \label{sec:mc}
After modelling the CV population, we perform Monte Carlo (MC) simulations to estimate the number of detectable nova remnants.
In each simulation, we first generate a mock CV sample with $N_{\rm CV}=10000$, whose WD masses and mass transfer rates follow the assumed distribution.
For each CV in the sample, the light curve of its nova remnant is determined with nearest WD mass and mass transfer rate in the model grid.
Then the observation time window, which is 13.5 years for {\it Chandra} observations toward the Galactic center and a negligible duration for radio and infrared observations, is randomly selected within a recurrence time cycle, defined as $P_{\rm rec}=M_{\rm acc}/\dot{M}$ following \citetalias{2005ApJ...623..398Y}.
We note that this simplified definition represent a lower limit on the recurrence time, since under certain conditions the WD can continue to gain mass \citep{2020AdSpR..66.1072H}.
For nova that is recurrent within the simulation, i.e., $P_{\rm rec}<1000~\rm yr$, the time window is selected within the simulation duration.
The observed luminosity is derived by averaging the synthetic light curve within the observation time window.
The MC simulation is performed for 1000 times, and we obtain the number of the detected novae.

\begin{figure}
    \centering
    \includegraphics[width=\linewidth]{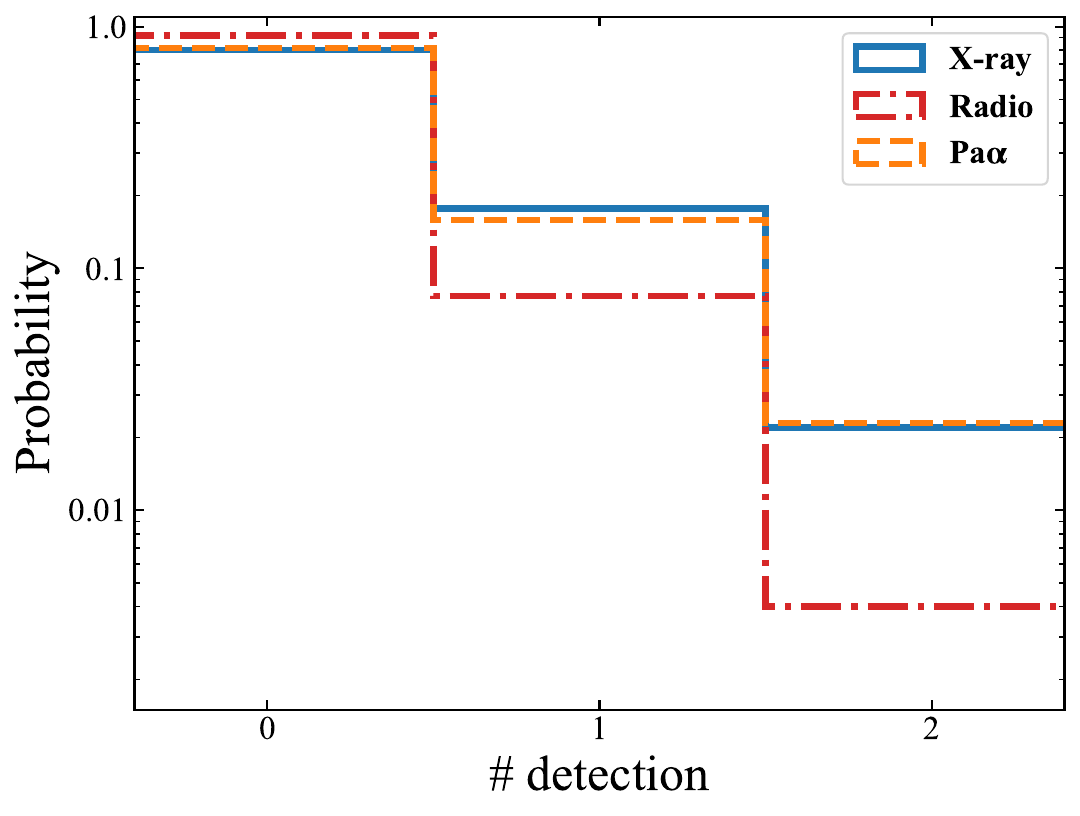}
    \caption{Multiwavelength detection probability distribution for the mock CV population ``field distribution". The number of detectable nova remnants (\#detection) is for $10^4$ CVs. \label{fig:multi_prob}}
\end{figure}

Figure \ref{fig:multi_prob} presents the simulated detection probabilities.
For the ``fiducial distribution", the probabilities of detecting at least one remnant among 10,000 CVs in the NSC are 20.0\%, 8.1\%, and 18.2\% for X-ray, radio, and Pa$\alpha$ emission, respectively.
These probabilities correspond to an average of $4.5\times10^4$, $1.2\times10^5$, and $4.9\times10^{4}$ CVs required to produce one detection, which is at least four times higher than the estimated CV population in the NSC.
While most nova models are detectable in radio and Pa$\alpha$ emission, the duration of their detectable windows is significantly shorter compared to X-rays. 
Furthermore, remnants that persist in radio and Pa$\alpha$ emissions tend to originate from novae with the most massive ejecta, which are associated with extremely long recurrence times.
The low detection probability and the distribution of the number of detectable remnants in Figure \ref{fig:multi_prob} suggest that detectable nova remnants would be rare in all three bands.
Despite uncertainties on the CV population modelling and synthetic luminosities, the rareness of detectable nova remnants appears realistic and aligns with the absence of observed nova outbursts or nova remnants in the NSC. 
To explore the impact of higher CV numbers as predicted by the simulations on star clusters and improved observational sensitivity, we consider a scenario with $N_{\rm CV}=30000$ and a deeper exposure by {\it Chandra} or future X-ray telescopes, such as AXIS \citep{2023SPIE12678E..1ER}, achieving a sensitivity of $L_{\rm det}=5\times10^{30}~\rm erg~s^{-1}$.
Under these assumptions, the detection probability increases significantly to 77.9\%, with an expectation of detecting 1–2 nova remnants in the NSC.

The probability of at least one detection by X-rays for the ``NSC distribution" is 5.2\%, which is four times lower than that for the ``field distribution".
The lower probability is because the detectable remnants from massive WDs tend to have low accretion rate and thus extremely long recurrence time of order $10^7~\rm year$ according to \citetalias{2005ApJ...623..398Y} model.
However, novae from the massive WDs would consequently have shorter recurrence time and higher possibility to be recurrent novae \citep{2018ApJ...860..110S, 2021MNRAS.504.6117K}.
Given the recurrence time from \citetalias{2005ApJ...623..398Y} model, the nova rate for ``NSC distribution" is $0.09~\rm yr^{-1}$, which is five times the nova rate for ``fiducial distribution", $0.018~\rm yr^{-1}$.
Therefore, although with lower probability of detectable remnants in X-rays, the novae for ``NSC distribution" could be potentially detected by the {\it Swift} monitoring of the GC for nearly 20 years \citep{2015JHEAp...7..137D}.
Moreover, the estimated nova rate is broadly consistent with \citet{2005ApJ...628..395T}, who predict nova rates of $0.033~\rm yr^{-1}$ and $0.011~\rm yr^{-1}$ for single WD mass of 1.0 and 0.6 $\rm M_\odot$, respectively.
The nova rate of $0.018~\rm yr^{-1}$ is also consistent with the Galactic rate of $50_{-23}^{+31}~\rm yr^{-1}$ \citep{2017ApJ...834..196S}, which corresponds to $0.025_{-0.012}^{+0.016}~\rm yr^{-1}$ when scaled to the NSC mass.
Despite this consistency, the nova rate in the Galactic center remains uncertain, owing to the potentially peculiar stellar and CV population compared to the Milky Way.

\section{Discussion}\label{sec:discussion}

\subsection{Where are nova remnant candidates?}\label{subsec:candidate}
\begin{figure}
	\centering
    \includegraphics[width=\linewidth]{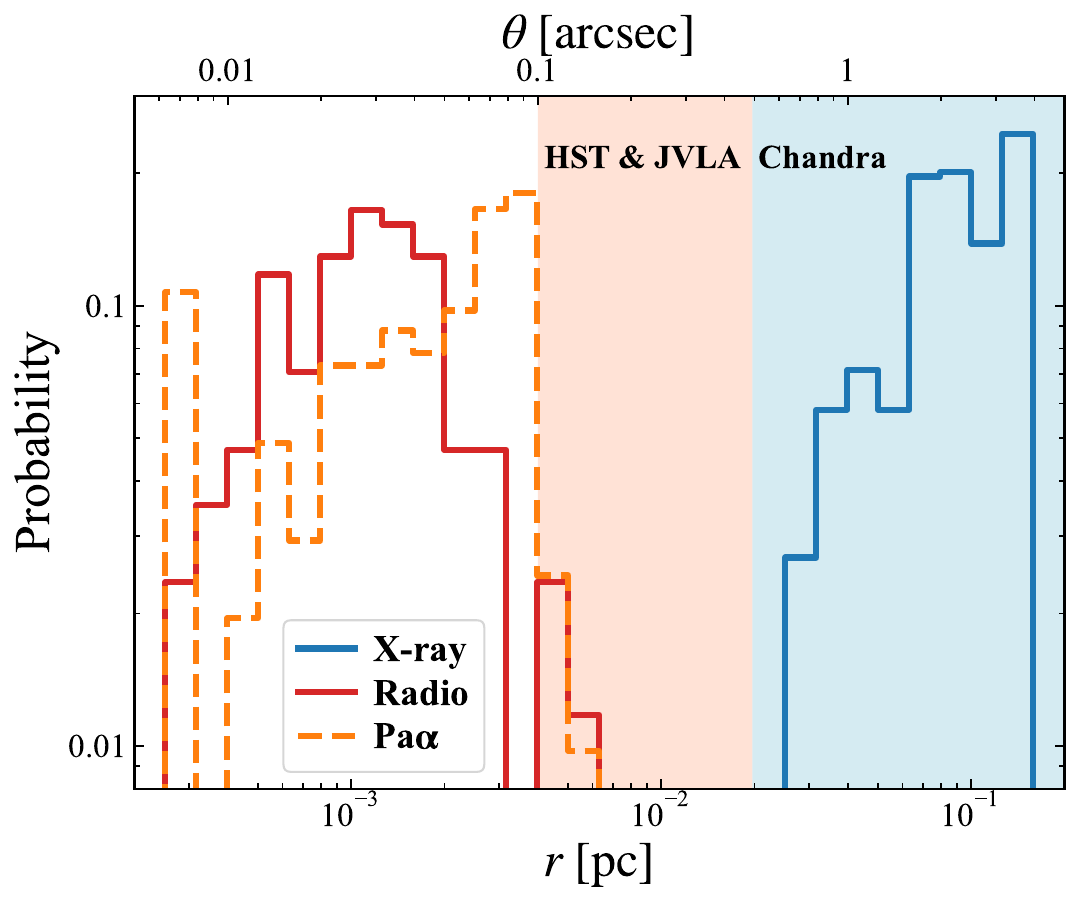}
    \caption{Probability distribution of multiwavelength radii of the detectable remnants from the Monte Carlo simulation. The shaded regions represent resolvable domain in X-ray (blue) and radio/Pa$\alpha$ (orange). \label{fig:resolve}}
\end{figure}

We have demonstrated the rarity of nova remnants in the GC for existing observations in Section \ref{sec:prob}. 
However, it is still worth addressing the observational properties of nova remnant candidates, in case some of them are already present in existed data.
In terms of morphology, the shell geometry could serve as a critical feature to distinguish them.
As illustrated in Figure \ref{fig:resolve}, the radii of detectable remnants are 0.5--4.0 arcsec in X-rays, and 0.01--0.1 arcsec in radio/Pa$\alpha$.
This discrepancy arises because radio and Pa$\alpha$ emission mostly originate from the photoionized ejecta in the early stages, whereas X-ray emission is dominated by the shocked ISM in the late stages.
We approximate the minimum resolvable radii for remnants to be 0.5 arcsec in X-rays, and 0.1 arcsec in radio/Pa$\alpha$.

The distribution of radii indicates that the X-ray remnants could be resolved, thanks to the superb angular resolution of {\it Chandra}, which would also help to mitigate contamination from X-ray emission of the central CV.
\citet{2018ApJS..235...26Z} cataloged 199 extended X-ray sources within the central 500 arcsec of the GC.
Although these sources are mainly clumps associated with Sgr A East and X-ray filaments, the potentially existed nova remnants might be among these sources, characterized by shell-like geometries.
Point sources could also be potential candidates, particularly if the nova remnants are compact and not well resolved as extended source.

In radio and Pa$\alpha$, the likelihood of resolving remnants is much lower, at around 3\%.
Consequently, the resolved GCCR sources in Figure. \ref{fig:radio_size} could hardly be associated with nova remnants.
Notably, approximately 1/3 of the GCCR sources in \citet{2020ApJ...905..173Z} are unresolved.
Additionally, about 40\% of the GCCR sources exhibit flat or rising spectra with spectral indices $-0.5<\alpha\le 1.65$, indicating a thermal origin for these radio sources.
Therefore, the thermal, unresolved GCCR sources, especially those manifested as variable or transient, could thus be potential candidates for nova remnants in radio band.
For Pa$\alpha$, the unresolved Pa$\alpha$ emitters in the GC are primarily determined to be young massive stars with strong stellar winds \citep{2011MNRAS.417..114D}.
Compared to these massive stars, CVs are much fainter in the infrared but can be much brighter in X-rays, with the exception of rare systems involving colliding-wind massive star binaries.
Therefore, Pa$\alpha$ sources that have an X-ray counterpart, i.e., the central CV, but lack a persistent broad-band infrared counterpart, could potentially be associated with nova remnants.

\subsection{X-ray measurement of metal abundances}
X-ray emission is an important tracer for ISM abundances in the Galactic center as the hot ISM usually stays on high ionization state and UV-optical emission is unreachable.
The X-ray spectrum of the nova remnant, brightened in X-rays by its own shock, can provide insights into elemental abundances of the ISM and/or nova ejecta.
The former is associated with the complex star formation history and anomalous initial mass function \citep{2010RvMP...82.3121G}, while the latter is highly relevant to the CV population that is suggested to contain a significant proportion of massive WDs in the NSC \citep{2015Natur.520..646P, 2019ApJ...882..164X, 2022MNRAS.511.5570H, 2022MNRAS.509.1175K}.
WD masses of these CVs, which are derived from surface temperature determined by X-ray spectral analysis, are estimated to be $1.0-1.2~\rm M_\odot$ if they are non-magnetic CVs \citep{2019ApJ...882..164X}.
This mass is above the natal mass of oxygen-neon (ONe) WD, a subclass of WD with different chemical makeup from common carbon-oxygen (CO) WDs.
Compared with CO WDs, nova ejecta from ONe WDs would have up to two orders of magnitude higher abundances of intermediate-mass elements (such as Ne, Mg, Si, S) \citep{1998ApJ...494..680J}. 
Since the UV-optical emission is extremely obscured, emission lines in hard X-rays from intermediate-mass elements would help discriminate between ONe and CO WDs.
And the mixing between ejecta and ISM makes the remnant characteristic of special elemental abundances dependent on the type of WDs.
However, in our simulation, the majority of X-ray emission comes from the shocked ISM while the ejecta has negligible contribution due to relatively low temperature $T\lesssim10^6~\rm K$.
In fact, mixing between the ejecta and ISM can still occur due to Rayleigh-Taylor instabilities and turbulent mixing in the remnant shell, which is ignored in our 1D hydrodynamical simulations.
Especially, the Rayleigh-Taylor mixing could strengthen when it happens close to the very Galactic center, under the gravitational potential of Sgr A* and the NSC. 

\subsection{Caveats}
Finally, we provide several caveats that could potentially influence the results of this work.
These include assumptions related to X-ray synthesis and physical properties of the ISM, as well as the inherent limitations of our simulations.

First, in the synthesis of X-ray emission, we assume an electron-ion temperature equilibrium, i.e., $T_e=T$.
However, this may not always be the case, particularly in shocked gas, where the majority of the energy is initially transferred to the ions.
Electrons are subsequently heated through elastic Coulomb collisions with the ions to establish temperature equilibrium.
At the time of the shock passage, the initial electron-to-ion temperature ratio $T_e/T$ ranges from $\sim0.1-1.0$, depending on Mach number $M$ of the shock \citep{2015A&A...579A..13V}.
The Mach number $M$ in our simulations generally falls in the range $\sim1-30$, where shocks in supernova remnants are observed to maintain $T_e/T$ close to 1 \citep{2023ApJ...949...50R}.
Therefore, the majority of our models would not significantly deviate from the temperature equilibrium, with exceptions in few cases involving extremely high ejecta velocities, such as $M \sim 80$ for model {\tt N48}.
With a reduced electron temperature, X-ray emission from the shocked gas would therefore be lower than our current estimation. 

Second, owing to the absence of radiation in our simulations, we adopt a simplified treatment in which the nova ejecta undergoes an early isothermal expansion according to \citet{2015ApJ...803...76C}.
In this framework, the duration of the isothermal phase $t_{\rm iso}$, the assumed ejecta temperature $T_{\rm ion}$, and the photon energy emitted by the post-eruption WD are critical for modeling the emission from the photoionized ejecta, particularly the radio and Pa$\alpha$ emission.
For instance, the detectable window for the Pa$\alpha$ emission shows a rough correlation with $t_{\rm iso}$, as the ejecta cool rapidly once they transition to adiabatic expansion after the isothermal phase.
Although our simulations broadly reproduce the observed radio and Pa$\alpha$ evolution (Figure \ref{fig:radio_compare} and \ref{fig:Pa_lc}), simulations with explicit WD radiation and a more extensive time-dependent ionization network would yield a more sophisticated and self-consistent model, albeit at substantially higher computational cost.

Third, in our simulations, we assume a single set of physical parameters to represent the ISM in the GC, i.e., $T_e=10^6~\rm K$ and $n_e=10~\rm cm^{-3}$.
However, there exists density and temperature gradients for the expanding and colliding stellar winds in the GC. 
The temperature and density could be significantly higher in the vicinity of Sgr A*, potentially making the remnant brighter in X-rays and increasing its detectability.
Additionally, the mass-segregation effect may lead to a concentration of massive objects, including CVs, in the inner region of the NSC.
Another limitation of our simulation setup is the assumption of a static ISM, which may not accurately reflect the environment of the GC.
In fact, nuclear outflow with velocities of several hundred $\rm km~s^{-1}$, driven by the stellar winds from dozens of Wolf-Rayet stars, may fill the inner parsecs of the GC.
Collisions between the outflow and nova ejecta could result in higher shock velocities, especially for novae with slow ejecta, such as those with $v_{\rm ej}\sim100~\rm km~s^{-1}$. 
Therefore, those slow novae interacting with the nuclear outflow near Sgr A* could still produce significant X-ray emission, whereas our simulations, which assume no such interaction, predict lower X-ray luminosities for these novae.

Finally, we employ a 1-D spherical grid in our simulations, which inherently neglects the clumpiness of the ejecta and other anisotropic effects.
These effects could induce a lower filling factor $f<1$ of the radiative plasma, as observed in phenomena such as the X-ray knots in the remnants of GK Per and DQ Her \citep{2005ApJ...627..933B,2003ApJ...594..428M} and optical clumps seen in other novae \citep{2008clno.book.....B}.
Multi-dimensional simulations could incorporate these anisotropic effects and the nuclear outflow to address this limitation.

\section{Conclusions} \label{sec:summary}
With the aim of investigating the detectability of nova remnants in the Galactic center, we have performed 1D hydrodynamical simulations of nova remnants evolving in the hot ISM with $T_e=10^{6}~\rm K$ and $n_e=10~\rm cm^{-3}$. A grid of 79 nova models from \citet{2005ApJ...623..398Y}, covering a broad parameter space, is adopted in the simulation. Our main results are as follows:
\begin{itemize}
	\item Nova ejecta initially undergo isothermal expansion ($\sim 10^4~\rm K$) before decelerating and oscillating at a certain radius with temperatures of $10^5-10^6~\rm K$. The ISM can be shocked and heated to $>10^7~\rm K$, depending on ejecta velocity.
	\item We estimate multi-wavelength emission from the nova remnants that can be revealed by present observations on the GC: thermal emission from the shocked ISM in hard X-rays, free-free emission in radio band, and Pa$\alpha$ recombination line in near-infrared. Of 79 models, 6, 44, and 51 are detectable by current {\it Chandra}, JVLA, and HST observations, respectively, demonstrating the potential to search for nova remnants in the GC. The detectable window ranges from weeks to more than one hundred years.
	\item To further estimate the possibility of detection, a CV population is assumed for the nuclear star cluster. Through Monte Carlo simulations, we estimate the possibilities of detecting the nova remnants to be 20.0\%, 8.1\%, and 18.2\% in X-rays, radio, and Pa$\alpha$.  Although nova remnants appear to be rare in the GC based on present observations, the depth and exceptional angular resolution of {\it Chandra} observations make X-rays the most promising probe for detecting and resolving these elusive remnants.
\end{itemize}
Future surveys of the GC, leveraging advanced facilities such as AXIS in X-rays \citep{2023SPIE12678E..1ER}, the Square Kilometer Array (SKA) \citep{2024arXiv240604022S} in radio, and JWST in infrared \citep{2023arXiv231011912S}, expected to achieve significantly improved sensitivity and resolution, greatly enhancing the likelihood of detecting and resolving nova remnants, and enabling a handful of detections.

\section*{Acknowledgements}

This work is supported by the National Key Research and Development Program of China (No. 2022YFF0503402), the National Natural Science Foundation of China (grant 12225302), the Fundamental Research Funds for the Central Universities (KG202502), and CNSA program D050102. The authors wish to thank the anonymous referee for helpful comments that improve this work, and Dr Yang Chen for helpful discussions.

\section*{Data Availability}
The data underlying this article will be shared on reasonable request to the corresponding author.



\bibliographystyle{mnras}
\bibliography{draft} 

\begin{thebibliography}{}
\makeatletter
\relax
\def\mn@urlcharsother{\let\do\@makeother \do\$\do\&\do\#\do\^\do\_\do\%\do\~}
\def\mn@doi{\begingroup\mn@urlcharsother \@ifnextchar [ {\mn@doi@} {\mn@doi@[]}}
\def\mn@doi@[#1]#2{\def\@tempa{#1}\ifx\@tempa\@empty \href {http://dx.doi.org/#2} {doi:#2}\else \href {http://dx.doi.org/#2} {#1}\fi \endgroup}
\def\mn@eprint#1#2{\mn@eprint@#1:#2::\@nil}
\def\mn@eprint@arXiv#1{\href {http://arxiv.org/abs/#1} {{\tt arXiv:#1}}}
\def\mn@eprint@dblp#1{\href {http://dblp.uni-trier.de/rec/bibtex/#1.xml} {dblp:#1}}
\def\mn@eprint@#1:#2:#3:#4\@nil{\def\@tempa {#1}\def\@tempb {#2}\def\@tempc {#3}\ifx \@tempc \@empty \let \@tempc \@tempb \let \@tempb \@tempa \fi \ifx \@tempb \@empty \def\@tempb {arXiv}\fi \@ifundefined {mn@eprint@\@tempb}{\@tempb:\@tempc}{\expandafter \expandafter \csname mn@eprint@\@tempb\endcsname \expandafter{\@tempc}}}

\bibitem[\protect\citeauthoryear{{Baganoff} et~al.,}{{Baganoff} et~al.}{2003}]{2003ApJ...591..891B}
{Baganoff} F.~K.,  et~al., 2003, \mn@doi [\apj] {10.1086/375145}, \href {https://ui.adsabs.harvard.edu/abs/2003ApJ...591..891B} {591, 891}

\bibitem[\protect\citeauthoryear{{Bailer-Jones}, {Rybizki}, {Fouesneau}, {Demleitner}  \& {Andrae}}{{Bailer-Jones} et~al.}{2021}]{2021AJ....161..147B}
{Bailer-Jones} C.~A.~L.,  {Rybizki} J.,  {Fouesneau} M.,  {Demleitner} M.,   {Andrae} R.,  2021, \mn@doi [\aj] {10.3847/1538-3881/abd806}, \href {https://ui.adsabs.harvard.edu/abs/2021AJ....161..147B} {161, 147}

\bibitem[\protect\citeauthoryear{{Balman}}{{Balman}}{2005}]{2005ApJ...627..933B}
{Balman} {\c{S}}.,  2005, \mn@doi [\apj] {10.1086/430392}, \href {https://ui.adsabs.harvard.edu/abs/2005ApJ...627..933B} {627, 933}

\bibitem[\protect\citeauthoryear{{Balman}}{{Balman}}{2006}]{2006AdSpR..38.2840B}
{Balman} {\c{S}}.,  2006, \mn@doi [Advances in Space Research] {10.1016/j.asr.2005.12.028}, \href {https://ui.adsabs.harvard.edu/abs/2006AdSpR..38.2840B} {38, 2840}

\bibitem[\protect\citeauthoryear{{Balman}}{{Balman}}{2010}]{2010MNRAS.404L..26B}
{Balman} {\c{S}}.,  2010, \mn@doi [\mnras] {10.1111/j.1745-3933.2010.00827.x}, \href {https://ui.adsabs.harvard.edu/abs/2010MNRAS.404L..26B} {404, L26}

\bibitem[\protect\citeauthoryear{{Balman}}{{Balman}}{2014}]{2014A&A...572A.114B}
{Balman} {\c{S}}.,  2014, \mn@doi [\aap] {10.1051/0004-6361/201424233}, \href {https://ui.adsabs.harvard.edu/abs/2014A&A...572A.114B} {572, A114}

\bibitem[\protect\citeauthoryear{{Balman}, {Orio}  \& {Luna}}{{Balman} et~al.}{2025}]{2025Univ...11..105B}
{Balman} {\c{S}}.,  {Orio} M.,   {Luna} G. J.~M.,  2025, \mn@doi [Universe] {10.3390/universe11040105}, \href {https://ui.adsabs.harvard.edu/abs/2025Univ...11..105B} {11, 105}

\bibitem[\protect\citeauthoryear{{Bao}, {Li}, {Cheng}  \& {Belloni}}{{Bao} et~al.}{2024}]{2024MNRAS.527.7173B}
{Bao} T.,  {Li} Z.,  {Cheng} Z.,   {Belloni} D.,  2024, \mn@doi [\mnras] {10.1093/mnras/stad3665}, \href {https://ui.adsabs.harvard.edu/abs/2024MNRAS.527.7173B} {527, 7173}

\bibitem[\protect\citeauthoryear{{Barna}, {Palou{\v{s}}}, {Ehlerov{\'a}}, {W{\"u}nsch}, {Morris}  \& {Vermot}}{{Barna} et~al.}{2022}]{2022MNRAS.510.5266B}
{Barna} B.,  {Palou{\v{s}}} J.,  {Ehlerov{\'a}} S.,  {W{\"u}nsch} R.,  {Morris} M.~R.,   {Vermot} P.,  2022, \mn@doi [\mnras] {10.1093/mnras/stab3723}, \href {https://ui.adsabs.harvard.edu/abs/2022MNRAS.510.5266B} {510, 5266}

\bibitem[\protect\citeauthoryear{{Belloni}, {Giersz}, {Rivera Sandoval}, {Askar}  \& {Cieciel{\r{a}}g}}{{Belloni} et~al.}{2019}]{2019MNRAS.483..315B}
{Belloni} D.,  {Giersz} M.,  {Rivera Sandoval} L.~E.,  {Askar} A.,   {Cieciel{\r{a}}g} P.,  2019, \mn@doi [\mnras] {10.1093/mnras/sty3097}, \href {https://ui.adsabs.harvard.edu/abs/2019MNRAS.483..315B} {483, 315}

\bibitem[\protect\citeauthoryear{{Bode} \& {Evans}}{{Bode} \& {Evans}}{2008}]{2008clno.book.....B}
{Bode} M.~F.,  {Evans} A.,  2008, {Classical Novae}.
 Vol. 43

\bibitem[\protect\citeauthoryear{{Bovy}}{{Bovy}}{2017}]{2017MNRAS.470.1360B}
{Bovy} J.,  2017, \mn@doi [\mnras] {10.1093/mnras/stx1277}, \href {https://ui.adsabs.harvard.edu/abs/2017MNRAS.470.1360B} {470, 1360}

\bibitem[\protect\citeauthoryear{{Canbay}, {Bilir}, {{\"O}zd{\"o}nmez}  \& {Ak}}{{Canbay} et~al.}{2023}]{2023AJ....165..163C}
{Canbay} R.,  {Bilir} S.,  {{\"O}zd{\"o}nmez} A.,   {Ak} T.,  2023, \mn@doi [\aj] {10.3847/1538-3881/acbead}, \href {https://ui.adsabs.harvard.edu/abs/2023AJ....165..163C} {165, 163}

\bibitem[\protect\citeauthoryear{{Chen}, {Woods}, {Yungelson}, {Gilfanov}  \& {Han}}{{Chen} et~al.}{2016}]{2016MNRAS.458.2916C}
{Chen} H.-L.,  {Woods} T.~E.,  {Yungelson} L.~R.,  {Gilfanov} M.,   {Han} Z.,  2016, \mn@doi [\mnras] {10.1093/mnras/stw458}, \href {https://ui.adsabs.harvard.edu/abs/2016MNRAS.458.2916C} {458, 2916}

\bibitem[\protect\citeauthoryear{{Chen}, {Woods}, {Yungelson}, {Piersanti}, {Gilfanov}  \& {Han}}{{Chen} et~al.}{2019}]{2019MNRAS.490.1678C}
{Chen} H.-L.,  {Woods} T.~E.,  {Yungelson} L.~R.,  {Piersanti} L.,  {Gilfanov} M.,   {Han} Z.,  2019, \mn@doi [\mnras] {10.1093/mnras/stz2644}, \href {https://ui.adsabs.harvard.edu/abs/2019MNRAS.490.1678C} {490, 1678}

\bibitem[\protect\citeauthoryear{{Chen} et~al.,}{{Chen} et~al.}{2023}]{2023ApJ...944...79C}
{Chen} Z.,  et~al., 2023, \mn@doi [\apj] {10.3847/1538-4357/aca8ad}, \href {https://ui.adsabs.harvard.edu/abs/2023ApJ...944...79C} {944, 79}

\bibitem[\protect\citeauthoryear{{Chomiuk}, {Metzger}  \& {Shen}}{{Chomiuk} et~al.}{2021a}]{2021ARA&A..59..391C}
{Chomiuk} L.,  {Metzger} B.~D.,   {Shen} K.~J.,  2021a, \mn@doi [\araa] {10.1146/annurev-astro-112420-114502}, \href {https://ui.adsabs.harvard.edu/abs/2021ARA&A..59..391C} {59, 391}

\bibitem[\protect\citeauthoryear{{Chomiuk} et~al.,}{{Chomiuk} et~al.}{2021b}]{2021ApJS..257...49C}
{Chomiuk} L.,  et~al., 2021b, \mn@doi [\apjs] {10.3847/1538-4365/ac24ab}, \href {https://ui.adsabs.harvard.edu/abs/2021ApJS..257...49C} {257, 49}

\bibitem[\protect\citeauthoryear{{Cunningham}, {Wolf}  \& {Bildsten}}{{Cunningham} et~al.}{2015}]{2015ApJ...803...76C}
{Cunningham} T.,  {Wolf} W.~M.,   {Bildsten} L.,  2015, \mn@doi [\apj] {10.1088/0004-637X/803/2/76}, \href {https://ui.adsabs.harvard.edu/abs/2015ApJ...803...76C} {803, 76}

\bibitem[\protect\citeauthoryear{{Curtin}, {Shafter}, {Pritchet}, {Neill}, {Kundu}  \& {Maccarone}}{{Curtin} et~al.}{2015}]{2015ApJ...811...34C}
{Curtin} C.,  {Shafter} A.~W.,  {Pritchet} C.~J.,  {Neill} J.~D.,  {Kundu} A.,   {Maccarone} T.~J.,  2015, \mn@doi [\apj] {10.1088/0004-637X/811/1/34}, \href {https://ui.adsabs.harvard.edu/abs/2015ApJ...811...34C} {811, 34}

\bibitem[\protect\citeauthoryear{{Das}, {Banerjee}, {Nandi}, {Ashok}  \& {Mondal}}{{Das} et~al.}{2015}]{2015MNRAS.447..806D}
{Das} R.,  {Banerjee} D. P.~K.,  {Nandi} A.,  {Ashok} N.~M.,   {Mondal} S.,  2015, \mn@doi [\mnras] {10.1093/mnras/stu2488}, \href {https://ui.adsabs.harvard.edu/abs/2015MNRAS.447..806D} {447, 806}

\bibitem[\protect\citeauthoryear{{De} et~al.,}{{De} et~al.}{2021}]{2021ApJ...912...19D}
{De} K.,  et~al., 2021, \mn@doi [\apj] {10.3847/1538-4357/abeb75}, \href {https://ui.adsabs.harvard.edu/abs/2021ApJ...912...19D} {912, 19}

\bibitem[\protect\citeauthoryear{{Degenaar}, {Wijnands}, {Miller}, {Reynolds}, {Kennea}  \& {Gehrels}}{{Degenaar} et~al.}{2015}]{2015JHEAp...7..137D}
{Degenaar} N.,  {Wijnands} R.,  {Miller} J.~M.,  {Reynolds} M.~T.,  {Kennea} J.,   {Gehrels} N.,  2015, \mn@doi [Journal of High Energy Astrophysics] {10.1016/j.jheap.2015.03.005}, \href {https://ui.adsabs.harvard.edu/abs/2015JHEAp...7..137D} {7, 137}

\bibitem[\protect\citeauthoryear{{Do} et~al.,}{{Do} et~al.}{2019}]{2019Sci...365..664D}
{Do} T.,  et~al., 2019, \mn@doi [Science] {10.1126/science.aav8137}, \href {https://ui.adsabs.harvard.edu/abs/2019Sci...365..664D} {365, 664}

\bibitem[\protect\citeauthoryear{{Dong} et~al.,}{{Dong} et~al.}{2011}]{2011MNRAS.417..114D}
{Dong} H.,  et~al., 2011, \mn@doi [\mnras] {10.1111/j.1365-2966.2011.19013.x}, \href {https://ui.adsabs.harvard.edu/abs/2011MNRAS.417..114D} {417, 114}

\bibitem[\protect\citeauthoryear{{Draine}}{{Draine}}{2011}]{2011piim.book.....D}
{Draine} B.~T.,  2011, {Physics of the Interstellar and Intergalactic Medium}

\bibitem[\protect\citeauthoryear{{Franckowiak}, {Jean}, {Wood}, {Cheung}  \& {Buson}}{{Franckowiak} et~al.}{2018}]{2018A&A...609A.120F}
{Franckowiak} A.,  {Jean} P.,  {Wood} M.,  {Cheung} C.~C.,   {Buson} S.,  2018, \mn@doi [\aap] {10.1051/0004-6361/201731516}, \href {https://ui.adsabs.harvard.edu/abs/2018A&A...609A.120F} {609, A120}

\bibitem[\protect\citeauthoryear{{GRAVITY Collaboration} et~al.,}{{GRAVITY Collaboration} et~al.}{2019}]{2019A&A...625L..10G}
{GRAVITY Collaboration} et~al., 2019, \mn@doi [\aap] {10.1051/0004-6361/201935656}, \href {https://ui.adsabs.harvard.edu/abs/2019A&A...625L..10G} {625, L10}

\bibitem[\protect\citeauthoryear{{Gaia Collaboration} et~al.,}{{Gaia Collaboration} et~al.}{2016}]{2016A&A...595A...1G}
{Gaia Collaboration} et~al., 2016, \mn@doi [\aap] {10.1051/0004-6361/201629272}, \href {https://ui.adsabs.harvard.edu/abs/2016A&A...595A...1G} {595, A1}

\bibitem[\protect\citeauthoryear{{Gaia Collaboration} et~al.,}{{Gaia Collaboration} et~al.}{2021}]{2021A&A...649A...1G}
{Gaia Collaboration} et~al., 2021, \mn@doi [\aap] {10.1051/0004-6361/202039657}, \href {https://ui.adsabs.harvard.edu/abs/2021A&A...649A...1G} {649, A1}

\bibitem[\protect\citeauthoryear{{Genzel}, {Eisenhauer}  \& {Gillessen}}{{Genzel} et~al.}{2010}]{2010RvMP...82.3121G}
{Genzel} R.,  {Eisenhauer} F.,   {Gillessen} S.,  2010, \mn@doi [Reviews of Modern Physics] {10.1103/RevModPhys.82.3121}, \href {https://ui.adsabs.harvard.edu/abs/2010RvMP...82.3121G} {82, 3121}

\bibitem[\protect\citeauthoryear{{Gillessen} et~al.,}{{Gillessen} et~al.}{2012}]{2012Natur.481...51G}
{Gillessen} S.,  et~al., 2012, \mn@doi [\nat] {10.1038/nature10652}, \href {https://ui.adsabs.harvard.edu/abs/2012Natur.481...51G} {481, 51}

\bibitem[\protect\citeauthoryear{{Gordon}, {Aydi}, {Page}, {Li}, {Chomiuk}, {Sokolovsky}, {Mukai}  \& {Seitz}}{{Gordon} et~al.}{2021}]{2021ApJ...910..134G}
{Gordon} A.~C.,  {Aydi} E.,  {Page} K.~L.,  {Li} K.-L.,  {Chomiuk} L.,  {Sokolovsky} K.~V.,  {Mukai} K.,   {Seitz} J.,  2021, \mn@doi [\apj] {10.3847/1538-4357/abe547}, \href {https://ui.adsabs.harvard.edu/abs/2021ApJ...910..134G} {910, 134}

\bibitem[\protect\citeauthoryear{{Harman} \& {O'Brien}}{{Harman} \& {O'Brien}}{2003}]{2003MNRAS.344.1219H}
{Harman} D.~J.,  {O'Brien} T.~J.,  2003, \mn@doi [\mnras] {10.1046/j.1365-8711.2003.06906.x}, \href {https://ui.adsabs.harvard.edu/abs/2003MNRAS.344.1219H} {344, 1219}

\bibitem[\protect\citeauthoryear{{Henze} et~al.,}{{Henze} et~al.}{2011}]{2011A&A...533A..52H}
{Henze} M.,  et~al., 2011, \mn@doi [\aap] {10.1051/0004-6361/201015887}, \href {https://ui.adsabs.harvard.edu/abs/2011A&A...533A..52H} {533, A52}

\bibitem[\protect\citeauthoryear{{Henze} et~al.,}{{Henze} et~al.}{2013}]{2013A&A...549A.120H}
{Henze} M.,  et~al., 2013, \mn@doi [\aap] {10.1051/0004-6361/201220196}, \href {https://ui.adsabs.harvard.edu/abs/2013A&A...549A.120H} {549, A120}

\bibitem[\protect\citeauthoryear{{Henze} et~al.,}{{Henze} et~al.}{2014}]{2014A&A...563A...2H}
{Henze} M.,  et~al., 2014, \mn@doi [\aap] {10.1051/0004-6361/201322426}, \href {https://ui.adsabs.harvard.edu/abs/2014A&A...563A...2H} {563, A2}

\bibitem[\protect\citeauthoryear{{Hillman} \& {Gerbi}}{{Hillman} \& {Gerbi}}{2022}]{2022MNRAS.511.5570H}
{Hillman} Y.,  {Gerbi} M.,  2022, \mn@doi [\mnras] {10.1093/mnras/stac432}, \href {https://ui.adsabs.harvard.edu/abs/2022MNRAS.511.5570H} {511, 5570}

\bibitem[\protect\citeauthoryear{{Hillman} \& {Kashi}}{{Hillman} \& {Kashi}}{2021}]{2021MNRAS.501..201H}
{Hillman} Y.,  {Kashi} A.,  2021, \mn@doi [\mnras] {10.1093/mnras/staa3600}, \href {https://ui.adsabs.harvard.edu/abs/2021MNRAS.501..201H} {501, 201}

\bibitem[\protect\citeauthoryear{{Hillman}, {Prialnik}, {Kovetz}  \& {Shara}}{{Hillman} et~al.}{2015}]{2015MNRAS.446.1924H}
{Hillman} Y.,  {Prialnik} D.,  {Kovetz} A.,   {Shara} M.~M.,  2015, \mn@doi [\mnras] {10.1093/mnras/stu2235}, \href {https://ui.adsabs.harvard.edu/abs/2015MNRAS.446.1924H} {446, 1924}

\bibitem[\protect\citeauthoryear{{Hillman}, {Prialnik}, {Kovetz}  \& {Shara}}{{Hillman} et~al.}{2016}]{2016ApJ...819..168H}
{Hillman} Y.,  {Prialnik} D.,  {Kovetz} A.,   {Shara} M.~M.,  2016, \mn@doi [\apj] {10.3847/0004-637X/819/2/168}, \href {https://ui.adsabs.harvard.edu/abs/2016ApJ...819..168H} {819, 168}

\bibitem[\protect\citeauthoryear{{Hillman}, {Shara}, {Prialnik}  \& {Kovetz}}{{Hillman} et~al.}{2020a}]{2020NatAs...4..886H}
{Hillman} Y.,  {Shara} M.~M.,  {Prialnik} D.,   {Kovetz} A.,  2020a, \mn@doi [Nature Astronomy] {10.1038/s41550-020-1062-y}, \href {https://ui.adsabs.harvard.edu/abs/2020NatAs...4..886H} {4, 886}

\bibitem[\protect\citeauthoryear{{Hillman}, {Shara}, {Prialnik}  \& {Kovetz}}{{Hillman} et~al.}{2020b}]{2020AdSpR..66.1072H}
{Hillman} Y.,  {Shara} M.,  {Prialnik} D.,   {Kovetz} A.,  2020b, \mn@doi [Advances in Space Research] {10.1016/j.asr.2019.08.029}, \href {https://ui.adsabs.harvard.edu/abs/2020AdSpR..66.1072H} {66, 1072}

\bibitem[\protect\citeauthoryear{{Hua}, {Li}, {Zhang}, {Chen}  \& {Morris}}{{Hua} et~al.}{2023}]{2023MNRAS.522..635H}
{Hua} Z.,  {Li} Z.,  {Zhang} M.,  {Chen} Z.,   {Morris} M.~R.,  2023, \mn@doi [\mnras] {10.1093/mnras/stad1025}, \href {https://ui.adsabs.harvard.edu/abs/2023MNRAS.522..635H} {522, 635}

\bibitem[\protect\citeauthoryear{{Jos{\'e}} \& {Hernanz}}{{Jos{\'e}} \& {Hernanz}}{1998}]{1998ApJ...494..680J}
{Jos{\'e}} J.,  {Hernanz} M.,  1998, \mn@doi [\apj] {10.1086/305244}, \href {https://ui.adsabs.harvard.edu/abs/1998ApJ...494..680J} {494, 680}

\bibitem[\protect\citeauthoryear{{Jos{\'e}}, {Garc{\'\i}a-Berro}, {Hernanz}  \& {Gil-Pons}}{{Jos{\'e}} et~al.}{2007}]{2007ApJ...662L.103J}
{Jos{\'e}} J.,  {Garc{\'\i}a-Berro} E.,  {Hernanz} M.,   {Gil-Pons} P.,  2007, \mn@doi [\apjl] {10.1086/519521}, \href {https://ui.adsabs.harvard.edu/abs/2007ApJ...662L.103J} {662, L103}

\bibitem[\protect\citeauthoryear{{Jos{\'e}}, {Shore}  \& {Casanova}}{{Jos{\'e}} et~al.}{2020}]{2020A&A...634A...5J}
{Jos{\'e}} J.,  {Shore} S.~N.,   {Casanova} J.,  2020, \mn@doi [\aap] {10.1051/0004-6361/201936893}, \href {https://ui.adsabs.harvard.edu/abs/2020A&A...634A...5J} {634, A5}

\bibitem[\protect\citeauthoryear{{Kawash} et~al.,}{{Kawash} et~al.}{2022}]{2022ApJ...937...64K}
{Kawash} A.,  et~al., 2022, \mn@doi [\apj] {10.3847/1538-4357/ac8d5e}, \href {https://ui.adsabs.harvard.edu/abs/2022ApJ...937...64K} {937, 64}

\bibitem[\protect\citeauthoryear{{Kemp}, {Karakas}, {Casey}, {Izzard}, {Ruiter}, {Agrawal}, {Broekgaarden}  \& {Temmink}}{{Kemp} et~al.}{2021}]{2021MNRAS.504.6117K}
{Kemp} A.~J.,  {Karakas} A.~I.,  {Casey} A.~R.,  {Izzard} R.~G.,  {Ruiter} A.~J.,  {Agrawal} P.,  {Broekgaarden} F.~S.,   {Temmink} K.~D.,  2021, \mn@doi [\mnras] {10.1093/mnras/stab1160}, \href {https://ui.adsabs.harvard.edu/abs/2021MNRAS.504.6117K} {504, 6117}

\bibitem[\protect\citeauthoryear{{Kemp}, {Karakas}, {Casey}, {Kobayashi}  \& {Izzard}}{{Kemp} et~al.}{2022}]{2022MNRAS.509.1175K}
{Kemp} A.~J.,  {Karakas} A.~I.,  {Casey} A.~R.,  {Kobayashi} C.,   {Izzard} R.~G.,  2022, \mn@doi [\mnras] {10.1093/mnras/stab3103}, \href {https://ui.adsabs.harvard.edu/abs/2022MNRAS.509.1175K} {509, 1175}

\bibitem[\protect\citeauthoryear{{Krautter} et~al.,}{{Krautter} et~al.}{2002}]{2002AJ....124.2888K}
{Krautter} J.,  et~al., 2002, \mn@doi [\aj] {10.1086/343833}, \href {https://ui.adsabs.harvard.edu/abs/2002AJ....124.2888K} {124, 2888}

\bibitem[\protect\citeauthoryear{{Lessing}, {Shara}, {Hounsell}, {Mandel}, {Feder}  \& {Sparks}}{{Lessing} et~al.}{2024}]{2024ApJ...973..144L}
{Lessing} A.~M.,  {Shara} M.~M.,  {Hounsell} R.,  {Mandel} S.,  {Feder} N.,   {Sparks} W.,  2024, \mn@doi [\apj] {10.3847/1538-4357/ad70b7}, \href {https://ui.adsabs.harvard.edu/abs/2024ApJ...973..144L} {973, 144}

\bibitem[\protect\citeauthoryear{{Li}, {Morris}  \& {Baganoff}}{{Li} et~al.}{2013}]{2013ApJ...779..154L}
{Li} Z.,  {Morris} M.~R.,   {Baganoff} F.~K.,  2013, \mn@doi [\apj] {10.1088/0004-637X/779/2/154}, \href {https://ui.adsabs.harvard.edu/abs/2013ApJ...779..154L} {779, 154}

\bibitem[\protect\citeauthoryear{{Lynch}, {Rudy}, {Mazuk}  \& {Puetter}}{{Lynch} et~al.}{2000}]{2000ApJ...541..791L}
{Lynch} D.~K.,  {Rudy} R.~J.,  {Mazuk} S.,   {Puetter} R.~C.,  2000, \mn@doi [\apj] {10.1086/309453}, \href {https://ui.adsabs.harvard.edu/abs/2000ApJ...541..791L} {541, 791}

\bibitem[\protect\citeauthoryear{{Mason}, {Shore}, {De Gennaro Aquino}, {Izzo}, {Page}  \& {Schwarz}}{{Mason} et~al.}{2018}]{2018ApJ...853...27M}
{Mason} E.,  {Shore} S.~N.,  {De Gennaro Aquino} I.,  {Izzo} L.,  {Page} K.,   {Schwarz} G.~J.,  2018, \mn@doi [\apj] {10.3847/1538-4357/aaa247}, \href {https://ui.adsabs.harvard.edu/abs/2018ApJ...853...27M} {853, 27}

\bibitem[\protect\citeauthoryear{{Metzger}, {Hasco{\"e}t}, {Vurm}, {Beloborodov}, {Chomiuk}, {Sokoloski}  \& {Nelson}}{{Metzger} et~al.}{2014}]{2014MNRAS.442..713M}
{Metzger} B.~D.,  {Hasco{\"e}t} R.,  {Vurm} I.,  {Beloborodov} A.~M.,  {Chomiuk} L.,  {Sokoloski} J.~L.,   {Nelson} T.,  2014, \mn@doi [\mnras] {10.1093/mnras/stu844}, \href {https://ui.adsabs.harvard.edu/abs/2014MNRAS.442..713M} {442, 713}

\bibitem[\protect\citeauthoryear{{Meyer} \& {Meyer-Hofmeister}}{{Meyer} \& {Meyer-Hofmeister}}{2012}]{2012A&A...546L...2M}
{Meyer} F.,  {Meyer-Hofmeister} E.,  2012, \mn@doi [\aap] {10.1051/0004-6361/201220145}, \href {https://ui.adsabs.harvard.edu/abs/2012A&A...546L...2M} {546, L2}

\bibitem[\protect\citeauthoryear{{Mignone}, {Bodo}, {Massaglia}, {Matsakos}, {Tesileanu}, {Zanni}  \& {Ferrari}}{{Mignone} et~al.}{2007}]{2007ApJS..170..228M}
{Mignone} A.,  {Bodo} G.,  {Massaglia} S.,  {Matsakos} T.,  {Tesileanu} O.,  {Zanni} C.,   {Ferrari} A.,  2007, \mn@doi [\apjs] {10.1086/513316}, \href {https://ui.adsabs.harvard.edu/abs/2007ApJS..170..228M} {170, 228}

\bibitem[\protect\citeauthoryear{{Montez}, {Luna}, {Mukai}, {Sokoloski}  \& {Kastner}}{{Montez} et~al.}{2022}]{2022ApJ...926..100M}
{Montez} R.,  {Luna} G.~J.~M.,  {Mukai} K.,  {Sokoloski} J.~L.,   {Kastner} J.~H.,  2022, \mn@doi [\apj] {10.3847/1538-4357/ac4583}, \href {https://ui.adsabs.harvard.edu/abs/2022ApJ...926..100M} {926, 100}

\bibitem[\protect\citeauthoryear{{Moraes} \& {Diaz}}{{Moraes} \& {Diaz}}{2009}]{2009AJ....138.1541M}
{Moraes} M.,  {Diaz} M.,  2009, \mn@doi [\aj] {10.1088/0004-6256/138/6/1541}, \href {https://ui.adsabs.harvard.edu/abs/2009AJ....138.1541M} {138, 1541}

\bibitem[\protect\citeauthoryear{{Mukai}, {Still}  \& {Ringwald}}{{Mukai} et~al.}{2003}]{2003ApJ...594..428M}
{Mukai} K.,  {Still} M.,   {Ringwald} F.~A.,  2003, \mn@doi [\apj] {10.1086/376752}, \href {https://ui.adsabs.harvard.edu/abs/2003ApJ...594..428M} {594, 428}

\bibitem[\protect\citeauthoryear{{Mukai}, {Orio}  \& {Della Valle}}{{Mukai} et~al.}{2008}]{2008ApJ...677.1248M}
{Mukai} K.,  {Orio} M.,   {Della Valle} M.,  2008, \mn@doi [\apj] {10.1086/529362}, \href {https://ui.adsabs.harvard.edu/abs/2008ApJ...677.1248M} {677, 1248}

\bibitem[\protect\citeauthoryear{{Muno}, {Pfahl}, {Baganoff}, {Brandt}, {Ghez}, {Lu}  \& {Morris}}{{Muno} et~al.}{2005}]{2005ApJ...622L.113M}
{Muno} M.~P.,  {Pfahl} E.,  {Baganoff} F.~K.,  {Brandt} W.~N.,  {Ghez} A.,  {Lu} J.,   {Morris} M.~R.,  2005, \mn@doi [\apjl] {10.1086/429721}, \href {https://ui.adsabs.harvard.edu/abs/2005ApJ...622L.113M} {622, L113}

\bibitem[\protect\citeauthoryear{{{\"O}zd{\"o}nmez}, {Ege}, {G{\"u}ver}  \& {Ak}}{{{\"O}zd{\"o}nmez} et~al.}{2018}]{2018MNRAS.476.4162O}
{{\"O}zd{\"o}nmez} A.,  {Ege} E.,  {G{\"u}ver} T.,   {Ak} T.,  2018, \mn@doi [\mnras] {10.1093/mnras/sty432}, \href {https://ui.adsabs.harvard.edu/abs/2018MNRAS.476.4162O} {476, 4162}

\bibitem[\protect\citeauthoryear{{Pala} et~al.,}{{Pala} et~al.}{2022}]{2022MNRAS.510.6110P}
{Pala} A.~F.,  et~al., 2022, \mn@doi [\mnras] {10.1093/mnras/stab3449}, \href {https://ui.adsabs.harvard.edu/abs/2022MNRAS.510.6110P} {510, 6110}

\bibitem[\protect\citeauthoryear{{Palou{\v{s}}}, {Ehlerov{\'a}}, {W{\"u}nsch}  \& {Morris}}{{Palou{\v{s}}} et~al.}{2020}]{2020A&A...644A..72P}
{Palou{\v{s}}} J.,  {Ehlerov{\'a}} S.,  {W{\"u}nsch} R.,   {Morris} M.~R.,  2020, \mn@doi [\aap] {10.1051/0004-6361/202038768}, \href {https://ui.adsabs.harvard.edu/abs/2020A&A...644A..72P} {644, A72}

\bibitem[\protect\citeauthoryear{{Panei}, {Althaus}  \& {Benvenuto}}{{Panei} et~al.}{2000}]{2000A&A...353..970P}
{Panei} J.~A.,  {Althaus} L.~G.,   {Benvenuto} O.~G.,  2000, \mn@doi [\aap] {10.48550/arXiv.astro-ph/9909499}, \href {https://ui.adsabs.harvard.edu/abs/2000A&A...353..970P} {353, 970}

\bibitem[\protect\citeauthoryear{{Perez} et~al.,}{{Perez} et~al.}{2015}]{2015Natur.520..646P}
{Perez} K.,  et~al., 2015, \mn@doi [\nat] {10.1038/nature14353}, \href {https://ui.adsabs.harvard.edu/abs/2015Natur.520..646P} {520, 646}

\bibitem[\protect\citeauthoryear{{Quataert}}{{Quataert}}{2004}]{2004ApJ...613..322Q}
{Quataert} E.,  2004, \mn@doi [\apj] {10.1086/422973}, \href {https://ui.adsabs.harvard.edu/abs/2004ApJ...613..322Q} {613, 322}

\bibitem[\protect\citeauthoryear{{Raj} et~al.,}{{Raj} et~al.}{2015}]{2015AJ....149..136R}
{Raj} A.,  et~al., 2015, \mn@doi [\aj] {10.1088/0004-6256/149/4/136}, \href {https://ui.adsabs.harvard.edu/abs/2015AJ....149..136R} {149, 136}

\bibitem[\protect\citeauthoryear{{Raymond} et~al.,}{{Raymond} et~al.}{2023}]{2023ApJ...949...50R}
{Raymond} J.~C.,  et~al., 2023, \mn@doi [\apj] {10.3847/1538-4357/acc528}, \href {https://ui.adsabs.harvard.edu/abs/2023ApJ...949...50R} {949, 50}

\bibitem[\protect\citeauthoryear{{Ressler}, {Quataert}  \& {Stone}}{{Ressler} et~al.}{2018}]{2018MNRAS.478.3544R}
{Ressler} S.~M.,  {Quataert} E.,   {Stone} J.~M.,  2018, \mn@doi [\mnras] {10.1093/mnras/sty1146}, \href {https://ui.adsabs.harvard.edu/abs/2018MNRAS.478.3544R} {478, 3544}

\bibitem[\protect\citeauthoryear{{Reynolds} et~al.,}{{Reynolds} et~al.}{2023}]{2023SPIE12678E..1ER}
{Reynolds} C.~S.,  et~al., 2023, in {Siegmund} O.~H.,  {Hoadley} K.,  eds,  Society of Photo-Optical Instrumentation Engineers (SPIE) Conference Series Vol. 12678, UV, X-Ray, and Gamma-Ray Space Instrumentation for Astronomy XXIII. p. 126781E (\mn@eprint {arXiv} {2311.00780}), \mn@doi{10.1117/12.2677468}

\bibitem[\protect\citeauthoryear{{Ribeiro}, {Darnley}, {Bode}, {Munari}, {Harman}, {Steele}  \& {Meaburn}}{{Ribeiro} et~al.}{2011}]{2011MNRAS.412.1701R}
{Ribeiro} V.~A.~R.~M.,  {Darnley} M.~J.,  {Bode} M.~F.,  {Munari} U.,  {Harman} D.~J.,  {Steele} I.~A.,   {Meaburn} J.,  2011, \mn@doi [\mnras] {10.1111/j.1365-2966.2010.18006.x}, \href {https://ui.adsabs.harvard.edu/abs/2011MNRAS.412.1701R} {412, 1701}

\bibitem[\protect\citeauthoryear{{Rupen}, {Mioduszewski}  \& {Sokoloski}}{{Rupen} et~al.}{2008}]{2008ApJ...688..559R}
{Rupen} M.~P.,  {Mioduszewski} A.~J.,   {Sokoloski} J.~L.,  2008, \mn@doi [\apj] {10.1086/525555}, \href {https://ui.adsabs.harvard.edu/abs/2008ApJ...688..559R} {688, 559}

\bibitem[\protect\citeauthoryear{{Schaal} \& {Springel}}{{Schaal} \& {Springel}}{2015}]{2015MNRAS.446.3992S}
{Schaal} K.,  {Springel} V.,  2015, \mn@doi [\mnras] {10.1093/mnras/stu2386}, \href {https://ui.adsabs.harvard.edu/abs/2015MNRAS.446.3992S} {446, 3992}

\bibitem[\protect\citeauthoryear{{Sch{\"o}del}, {Feldmeier}, {Kunneriath}, {Stolovy}, {Neumayer}, {Amaro-Seoane}  \& {Nishiyama}}{{Sch{\"o}del} et~al.}{2014}]{2014A&A...566A..47S}
{Sch{\"o}del} R.,  {Feldmeier} A.,  {Kunneriath} D.,  {Stolovy} S.,  {Neumayer} N.,  {Amaro-Seoane} P.,   {Nishiyama} S.,  2014, \mn@doi [\aap] {10.1051/0004-6361/201423481}, \href {https://ui.adsabs.harvard.edu/abs/2014A&A...566A..47S} {566, A47}

\bibitem[\protect\citeauthoryear{{Schoedel} et~al.,}{{Schoedel} et~al.}{2023}]{2023arXiv231011912S}
{Schoedel} R.,  et~al., 2023, \mn@doi [arXiv e-prints] {10.48550/arXiv.2310.11912}, \href {https://ui.adsabs.harvard.edu/abs/2023arXiv231011912S} {p. arXiv:2310.11912}

\bibitem[\protect\citeauthoryear{{Schoedel} et~al.,}{{Schoedel} et~al.}{2024}]{2024arXiv240604022S}
{Schoedel} R.,  et~al., 2024, \mn@doi [arXiv e-prints] {10.48550/arXiv.2406.04022}, \href {https://ui.adsabs.harvard.edu/abs/2024arXiv240604022S} {p. arXiv:2406.04022}

\bibitem[\protect\citeauthoryear{{Shafter}}{{Shafter}}{2017}]{2017ApJ...834..196S}
{Shafter} A.~W.,  2017, \mn@doi [\apj] {10.3847/1538-4357/834/2/196}, \href {https://ui.adsabs.harvard.edu/abs/2017ApJ...834..196S} {834, 196}

\bibitem[\protect\citeauthoryear{{Shara}, {Prialnik}, {Hillman}  \& {Kovetz}}{{Shara} et~al.}{2018}]{2018ApJ...860..110S}
{Shara} M.~M.,  {Prialnik} D.,  {Hillman} Y.,   {Kovetz} A.,  2018, \mn@doi [\apj] {10.3847/1538-4357/aabfbd}, \href {https://ui.adsabs.harvard.edu/abs/2018ApJ...860..110S} {860, 110}

\bibitem[\protect\citeauthoryear{{Shen} \& {Bildsten}}{{Shen} \& {Bildsten}}{2009}]{2009ApJ...692..324S}
{Shen} K.~J.,  {Bildsten} L.,  2009, \mn@doi [\apj] {10.1088/0004-637X/692/1/324}, \href {https://ui.adsabs.harvard.edu/abs/2009ApJ...692..324S} {692, 324}

\bibitem[\protect\citeauthoryear{{Starrfield}, {Bose}, {Iliadis}, {Hix}, {Woodward}  \& {Wagner}}{{Starrfield} et~al.}{2020}]{2020ApJ...895...70S}
{Starrfield} S.,  {Bose} M.,  {Iliadis} C.,  {Hix} W.~R.,  {Woodward} C.~E.,   {Wagner} R.~M.,  2020, \mn@doi [\apj] {10.3847/1538-4357/ab8d23}, \href {https://ui.adsabs.harvard.edu/abs/2020ApJ...895...70S} {895, 70}

\bibitem[\protect\citeauthoryear{{Starrfield}, {Bose}, {Iliadis}, {Hix}, {Woodward}  \& {Wagner}}{{Starrfield} et~al.}{2024}]{2024ApJ...962..191S}
{Starrfield} S.,  {Bose} M.,  {Iliadis} C.,  {Hix} W.~R.,  {Woodward} C.~E.,   {Wagner} R.~M.,  2024, \mn@doi [\apj] {10.3847/1538-4357/ad1836}, \href {https://ui.adsabs.harvard.edu/abs/2024ApJ...962..191S} {962, 191}

\bibitem[\protect\citeauthoryear{{Storey} \& {Hummer}}{{Storey} \& {Hummer}}{1995}]{1995MNRAS.272...41S}
{Storey} P.~J.,  {Hummer} D.~G.,  1995, \mn@doi [\mnras] {10.1093/mnras/272.1.41}, \href {https://ui.adsabs.harvard.edu/abs/1995MNRAS.272...41S} {272, 41}

\bibitem[\protect\citeauthoryear{{Tappert}, {Vogt}, {Ederoclite}, {Schmidtobreick}, {Vu{\v{c}}kovi{\'c}}  \& {Becegato}}{{Tappert} et~al.}{2020}]{2020A&A...641A.122T}
{Tappert} C.,  {Vogt} N.,  {Ederoclite} A.,  {Schmidtobreick} L.,  {Vu{\v{c}}kovi{\'c}} M.,   {Becegato} L.~L.,  2020, \mn@doi [\aap] {10.1051/0004-6361/202037913}, \href {https://ui.adsabs.harvard.edu/abs/2020A&A...641A.122T} {641, A122}

\bibitem[\protect\citeauthoryear{{Te{\c{s}}ileanu}, {Mignone}  \& {Massaglia}}{{Te{\c{s}}ileanu} et~al.}{2008}]{2008A&A...488..429T}
{Te{\c{s}}ileanu} O.,  {Mignone} A.,   {Massaglia} S.,  2008, \mn@doi [\aap] {10.1051/0004-6361:200809461}, \href {https://ui.adsabs.harvard.edu/abs/2008A&A...488..429T} {488, 429}

\bibitem[\protect\citeauthoryear{{Toal{\'a}}, {Guerrero}, {Santamar{\'\i}a}, {Ramos-Larios}  \& {Sabin}}{{Toal{\'a}} et~al.}{2020}]{2020MNRAS.495.4372T}
{Toal{\'a}} J.~A.,  {Guerrero} M.~A.,  {Santamar{\'\i}a} E.,  {Ramos-Larios} G.,   {Sabin} L.,  2020, \mn@doi [\mnras] {10.1093/mnras/staa1502}, \href {https://ui.adsabs.harvard.edu/abs/2020MNRAS.495.4372T} {495, 4372}

\bibitem[\protect\citeauthoryear{{Townsley} \& {Bildsten}}{{Townsley} \& {Bildsten}}{2005}]{2005ApJ...628..395T}
{Townsley} D.~M.,  {Bildsten} L.,  2005, \mn@doi [\apj] {10.1086/430594}, \href {https://ui.adsabs.harvard.edu/abs/2005ApJ...628..395T} {628, 395}

\bibitem[\protect\citeauthoryear{{Vathachira}, {Hillman}  \& {Kashi}}{{Vathachira} et~al.}{2024}]{2024MNRAS.527.4806V}
{Vathachira} I.~B.,  {Hillman} Y.,   {Kashi} A.,  2024, \mn@doi [\mnras] {10.1093/mnras/stad3507}, \href {https://ui.adsabs.harvard.edu/abs/2024MNRAS.527.4806V} {527, 4806}

\bibitem[\protect\citeauthoryear{{Vink}, {Broersen}, {Bykov}  \& {Gabici}}{{Vink} et~al.}{2015}]{2015A&A...579A..13V}
{Vink} J.,  {Broersen} S.,  {Bykov} A.,   {Gabici} S.,  2015, \mn@doi [\aap] {10.1051/0004-6361/201424612}, \href {https://ui.adsabs.harvard.edu/abs/2015A&A...579A..13V} {579, A13}

\bibitem[\protect\citeauthoryear{{Wang} et~al.,}{{Wang} et~al.}{2010}]{2010MNRAS.402..895W}
{Wang} Q.~D.,  et~al., 2010, \mn@doi [\mnras] {10.1111/j.1365-2966.2009.15973.x}, \href {https://ui.adsabs.harvard.edu/abs/2010MNRAS.402..895W} {402, 895}

\bibitem[\protect\citeauthoryear{{Wolf}, {Bildsten}, {Brooks}  \& {Paxton}}{{Wolf} et~al.}{2013}]{2013ApJ...777..136W}
{Wolf} W.~M.,  {Bildsten} L.,  {Brooks} J.,   {Paxton} B.,  2013, \mn@doi [\apj] {10.1088/0004-637X/777/2/136}, \href {https://ui.adsabs.harvard.edu/abs/2013ApJ...777..136W} {777, 136}

\bibitem[\protect\citeauthoryear{{Xu}, {Wang}  \& {Li}}{{Xu} et~al.}{2016}]{2016ApJ...818..136X}
{Xu} X.-j.,  {Wang} Q.~D.,   {Li} X.-D.,  2016, \mn@doi [\apj] {10.3847/0004-637X/818/2/136}, \href {https://ui.adsabs.harvard.edu/abs/2016ApJ...818..136X} {818, 136}

\bibitem[\protect\citeauthoryear{{Xu}, {Li}, {Zhu}, {Cheng}, {Li}  \& {Yu}}{{Xu} et~al.}{2019}]{2019ApJ...882..164X}
{Xu} X.-j.,  {Li} Z.,  {Zhu} Z.,  {Cheng} Z.,  {Li} X.-d.,   {Yu} Z.-l.,  2019, \mn@doi [\apj] {10.3847/1538-4357/ab32df}, \href {https://ui.adsabs.harvard.edu/abs/2019ApJ...882..164X} {882, 164}

\bibitem[\protect\citeauthoryear{{Yalinewich}, {Piran}  \& {Sari}}{{Yalinewich} et~al.}{2017}]{2017ApJ...838...12Y}
{Yalinewich} A.,  {Piran} T.,   {Sari} R.,  2017, \mn@doi [\apj] {10.3847/1538-4357/aa5d0f}, \href {https://ui.adsabs.harvard.edu/abs/2017ApJ...838...12Y} {838, 12}

\bibitem[\protect\citeauthoryear{{Yaron}, {Prialnik}, {Shara}  \& {Kovetz}}{{Yaron} et~al.}{2005}]{2005ApJ...623..398Y}
{Yaron} O.,  {Prialnik} D.,  {Shara} M.~M.,   {Kovetz} A.,  2005, \mn@doi [\apj] {10.1086/428435}, \href {https://ui.adsabs.harvard.edu/abs/2005ApJ...623..398Y} {623, 398}

\bibitem[\protect\citeauthoryear{{Ye}, {Kremer}, {Rodriguez}, {Rui}, {Weatherford}, {Chatterjee}, {Fragione}  \& {Rasio}}{{Ye} et~al.}{2022}]{2022ApJ...931...84Y}
{Ye} C.~S.,  {Kremer} K.,  {Rodriguez} C.~L.,  {Rui} N.~Z.,  {Weatherford} N.~C.,  {Chatterjee} S.,  {Fragione} G.,   {Rasio} F.~A.,  2022, \mn@doi [\apj] {10.3847/1538-4357/ac5b0b}, \href {https://ui.adsabs.harvard.edu/abs/2022ApJ...931...84Y} {931, 84}

\bibitem[\protect\citeauthoryear{{Yu}, {Xu}, {Li}, {Bao}, {Li}, {Xing}  \& {Shen}}{{Yu} et~al.}{2018}]{2018ApJ...853..182Y}
{Yu} Z.-l.,  {Xu} X.-j.,  {Li} X.-D.,  {Bao} T.,  {Li} Y.-x.,  {Xing} Y.-c.,   {Shen} Y.-f.,  2018, \mn@doi [\apj] {10.3847/1538-4357/aaa47d}, \href {https://ui.adsabs.harvard.edu/abs/2018ApJ...853..182Y} {853, 182}

\bibitem[\protect\citeauthoryear{{Zhao}, {Morris}  \& {Goss}}{{Zhao} et~al.}{2020}]{2020ApJ...905..173Z}
{Zhao} J.-H.,  {Morris} M.~R.,   {Goss} W.~M.,  2020, \mn@doi [\apj] {10.3847/1538-4357/abc75e}, \href {https://ui.adsabs.harvard.edu/abs/2020ApJ...905..173Z} {905, 173}

\bibitem[\protect\citeauthoryear{{Zhao}, {Morris}  \& {Goss}}{{Zhao} et~al.}{2022}]{2022ApJ...927L...6Z}
{Zhao} J.-H.,  {Morris} M.~R.,   {Goss} W.~M.,  2022, \mn@doi [\apjl] {10.3847/2041-8213/ac54be}, \href {https://ui.adsabs.harvard.edu/abs/2022ApJ...927L...6Z} {927, L6}

\bibitem[\protect\citeauthoryear{{Zhu}, {Li}  \& {Morris}}{{Zhu} et~al.}{2018}]{2018ApJS..235...26Z}
{Zhu} Z.,  {Li} Z.,   {Morris} M.~R.,  2018, \mn@doi [\apjs] {10.3847/1538-4365/aab14f}, \href {https://ui.adsabs.harvard.edu/abs/2018ApJS..235...26Z} {235, 26}

\bibitem[\protect\citeauthoryear{{Zhu}, {Li}, {Morris}, {Zhang}  \& {Liu}}{{Zhu} et~al.}{2019}]{2019ApJ...875...44Z}
{Zhu} Z.,  {Li} Z.,  {Morris} M.~R.,  {Zhang} S.,   {Liu} S.,  2019, \mn@doi [\apj] {10.3847/1538-4357/ab0e05}, \href {https://ui.adsabs.harvard.edu/abs/2019ApJ...875...44Z} {875, 44}

\bibitem[\protect\citeauthoryear{{Zuckerman}, {De}, {Eilers}, {Meisner}  \& {Panagiotou}}{{Zuckerman} et~al.}{2023}]{2023MNRAS.523.3555Z}
{Zuckerman} L.,  {De} K.,  {Eilers} A.-C.,  {Meisner} A.~M.,   {Panagiotou} C.,  2023, \mn@doi [\mnras] {10.1093/mnras/stad1625}, \href {https://ui.adsabs.harvard.edu/abs/2023MNRAS.523.3555Z} {523, 3555}

\makeatother
\end{thebibliography}




\appendix
\section{Numerical Convergence}\label{sec:resolution}
\begin{figure*}
	\centering
    \includegraphics[width=\linewidth]{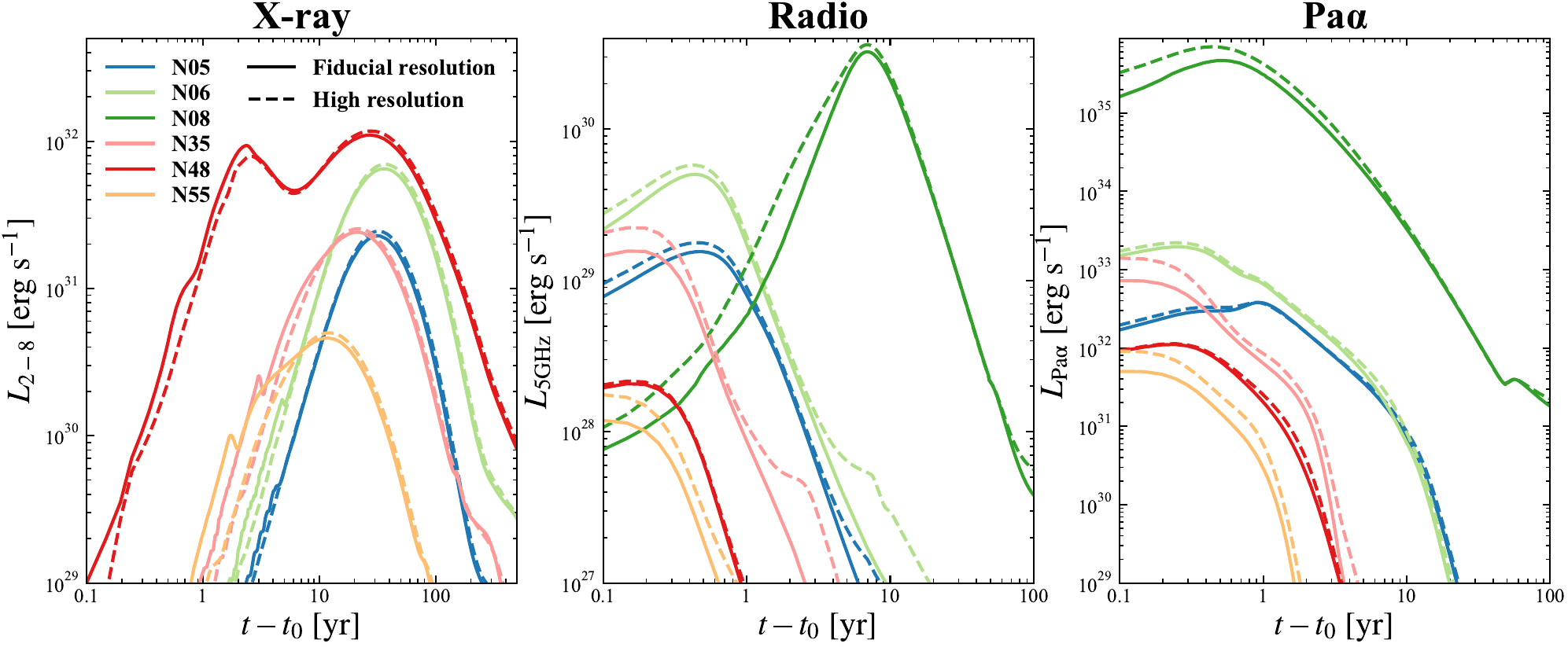}
    \caption{Comparison of synthetic light curves from simulations with different numerical resolutions. The ``fiducial resolution" (solid curves) corresponds to our fiducial simulations while the ``high resolution" (dashed curves) refers to simulations with five times higher spatial resolution. From left to right, the panels show light curves for the X-ray band (2--8 keV), the radio band (5 GHz), and the Pa$\alpha$ line, respectively. \label{fig:resolution}}
\end{figure*}

To test the numerical convergence of our simulations,  we further perform high-resolution simulations for six representative models (Table \ref{tab:resolution}), which span the parameter space in both ejecta velocity and ejecta mass.
The high-resolution simulations adopt a grid with 102400 cells, corresponding to a resolution five times higher than the fiducial simulations described in Section \ref{sec:hd}.
As shown in Figure \ref{fig:resolution}, the high-resolution simulations exhibit striking resemblance to the fiducial simulations in all three bands, except for the slightly enhanced radio and Pa$\alpha$ emission in the early stage.
Consistently, we also find almost identical detectable wind (Table \ref{tab:resolution}) that is critical for estimating the detection probability.
Therefore, we conclude that the present numerical resolution warrants a robust estimate of the detection probability of nova remnants, which is the focus of this work.

\begin{table*}
\centering
\begin{threeparttable}
\caption{Comparison between simulations with low and high numerical resolution \label{tab:resolution}}
\label{tab:resolution}
\renewcommand{\arraystretch}{1.0}
	{\footnotesize
	\centering
	\begin{tabular}{lcccccccc}	
	\hline
	Model & $M_{\rm ej}$ & $v_{\rm ej}$ & $L^{\rm peak}_{\rm X}$ & $t^{\rm det}_{\rm X}$ & $L^{\rm peak}_{\rm R}$ & $t^{\rm det}_{\rm R}$ & $L^{\rm peak}_{\rm Pa\alpha}$ & $t^{\rm det}_{\rm Pa\alpha}$ \\
	     &  $(\rm M_\odot)$  & $(\rm km~s^{-1})$ & $(\rm erg~s^{-1})$ & (yr) & $(\rm erg~s^{-1})$ & (yr) & $(\rm erg~s^{-1})$ & (yr) \\
	(1) & (2) & (3) & (4) & (5) & (6) & (7) & (8) & (9)\\
	 \hline
N05    & 1.6E-04 & 2150 & 2.3E+31 & 50.13    & 1.6E+29 & 1.61     & 4.2E+32 & 4.15     \\
N05HR  & 1.6E-04 & 2150 & 2.5E+31 & 54.15    & 1.9E+29 & 1.66     & 4.3E+32 & 4.15     \\
	 \hline
N06   & 2.8E-04 & 2650 & 6.5E+31 & 107.82   & 5.8E+29 & 1.95     & 2.6E+33 & 3.67     \\
N06HR & 2.8E-04 & 2650 & 7.0E+31 & 112.62   & 6.6E+29 & 2.05     & 2.9E+33 & 3.71     \\
	\hline
N08 & 6.7E-04 & 216 & 1.5E+25 & $\cdots$ & 3.3E+30 & 43.96    & 5.1E+35 & 37.38    \\
N08HR & 6.7E-04 & 216 & 1.5E+25 & $\cdots$ & 3.6E+30 & 44.24    & 7.7E+35 & 37.64    \\
	\hline
N35 & 5.5E-05 & 3490 & 2.4E+31 & 44.22    & 3.4E+29 & 0.49     & 2.8E+33 & 1.37     \\
N35HR & 5.5E-05 & 3490 & 2.5E+31 & 46.22    & 5.4E+29 & 0.54     & 5.3E+33 & 1.47     \\
	\hline
N48 & 5.3E-05 & 8267 & 1.1E+32 & 165.03   & 3.8E+28 & 0.15     & 1.6E+32 & 0.73     \\
N48HR & 5.3E-05 & 8267 & 1.2E+32 & 173.63   & 4.0E+28 & 0.15     & 1.7E+32 & 0.73     \\
	\hline
N55 & 9.9E-06 & 3668 & 4.6E+30 & $\cdots$ & 4.7E+28 & 0.05     & 1.7E+32 & 0.44     \\
N55HR & 9.9E-06 & 3668 & 5.0E+30 & $\cdots$ & 6.9E+28 & 0.09     & 3.6E+32 & 0.64     \\
	\hline
	\end{tabular}
	}
	\begin{tablenotes}
	{\footnotesize
	\item[]{\bf Notes.} The column definitions are the same as in Table \ref{tab:sum}. The model name ``N*HR" refer to simulations with a numerical resolution five times higher than the fiducial simulations, named ``N*" listed in Table \ref{tab:sum}.}
	\end{tablenotes}
	\end{threeparttable}
\end{table*}

%


\bsp	
\label{lastpage}
\end{document}